\documentclass[twocolumn]{svjour3}         
\smartqed  
\usepackage{hyperref}
\usepackage{graphicx}
\usepackage{lineno}
\usepackage[english]{babel}
\usepackage{epstopdf}
\usepackage{amsmath}
\usepackage{algorithm}
\usepackage{algpseudocode,algorithmicx}
\usepackage{graphicx}
\usepackage{float}
\usepackage{amssymb}
\usepackage{multirow}
\usepackage{caption}
\usepackage{color,ctable}
\usepackage[mathcal]{eucal}
\usepackage{times}
\usepackage[authoryear,numbers]{natbib}
\usepackage{amsmath,amsfonts,amssymb}
\usepackage[labelfont=bf, labelsep={space}]{caption}
\DeclareMathOperator{\E}{\mathbb{E}}
\begin{document}
\title{Parallel Tempering via Simulated Tempering Without Normalizing Constants\thanks{Acknowledgments: This work was partially funded by  Collaborative Research and Development (CRD) Grant involving the company FPInnovations. The authors would like to thank Dr. Luke Bornn, Dr. Derek Bingham, Dr. Liangliang Wang, Dr. Russell Steele and Dr. Mark Girolami for the constructive discussions, and Dr. Michael Jack Davis for proof reading the manuscript. The authors would like to thank the anonymous reviewer for their constructive remarks.}
}


\author{Biljana Jonoska Stojkova  \and
        David A. Campbell 
}


\institute{F. Author \at
              Department of Statistics \\
              University Of British Columbia \\
              3182 Earth Sciences Building, 2207 Main Mall \\
               Vancouver, BC Canada V6T 1Z4\\
              Tel: +1-604-822-2479\\ 
              \email{b.stojkova@stat.ubc.ca}  
           \and
           S. Author \at
             Department of Statistics and Actuarial Science\\ 
             Simon Fraser University \\
              8888 University Drive \\
              Burnaby, BC Canada, V5A 1S6 \\
              \email{dac5@sfu.ca}  
}

\date{Received: date / Accepted: date}

\maketitle

\begin{abstract}
In this paper we develop 
a new general Bayesian \\
methodology that 
simultaneously estimates 
parameters of interest and 
the marginal likelihood of the model. 
The proposed methodology builds on 
Simulated Tempering, which is a powerful algorithm that enables sampling from multi-modal distributions.  
However, Simulated Tempering comes with the practical limitation of needing to specify a prior for the temperature along a chosen discretization schedule that will allow calculation of normalizing constants at each temperature.  Our proposed model defines the prior for 
the temperature so as to remove the need for calculating normalizing constants at each temperature and thereby enables a continuous temperature schedule, while preserving the sampling efficiency of the Simulated Tempering algorithm.  
\\ The resulting algorithm simultaneously estimates parameters while estimating marginal likelihoods through thermo-\\ 
dynamic integration.  
We illustrate the applicability of the new algorithm to different examples involving mixture models of Gaussian distributions and ordinary differential equation models.  
\end{abstract}

\section{Introduction}
Basic random walk Markov Chain Monte Carlo (MCMC) methods are inefficient when faced with multi-modal posterior distributions, especially when modes are isolated by large gaps of low probability. Simulated Tempering \citep{MariParisi,ThompGeyer} and Parallel Tempering \citep{swendsen1986replica,Geyer,Hukushima}, are MCMC variants designed to ease the challenges of multi-modal distributions by incorporating an auxiliary temperature parameter to overcome the prohibitively low probability regions which otherwise trap samplers in local modes \citep{zhang2008comparison}. Sampling occurs at different temperatures, balancing short distance within-mode steps and longer distance between-mode steps. 

Both Parallel Tempering (PT) and Simulated Tempering (ST) define a sequence of distributions for the vector of data $\boldsymbol{Y} \in \mathbb{R}^{N}$ and parameter $\boldsymbol{\theta }\in \mathbb{R}^{d}$, indexed by an inverse temperature $\tau \in [0,1]$, along a path between the prior and the target distribution  usually defined by \[ P(\boldsymbol{\theta} \mid \tau,\boldsymbol{Y}) \propto P(\boldsymbol{Y} \mid  \boldsymbol{\theta},\tau)P(\boldsymbol{\theta})=P(\boldsymbol{Y} \mid  \boldsymbol{\theta})^{\tau}P(\boldsymbol{\theta}). \] At the extremes of the distribution sequence, $P(\boldsymbol{\theta} \mid \tau=0,\boldsymbol{Y})=P(\boldsymbol{\theta})$ is the prior, and $P(\boldsymbol{\theta} \mid \tau=1,\boldsymbol{Y})\propto P(\boldsymbol{Y} \mid  \boldsymbol{\theta})P(\boldsymbol{\theta})$ is the usual target  distribution.  Consequently, when $\tau=0$ it is easy for the sampler to explore the parameter space, but at $\tau=1$ the sampler explores the more challenging target distribution. Markov Chains in both PT and ST therefore avoid becoming trapped in a single mode.  In PT, $\text{T}$ independent chains are run at different temperatures where information about $\boldsymbol{\theta}$ is allowed to flow between them.  ST samples the system at different temperatures using a single chain by augmenting the state space with the temperature parameter as a dynamic variable. In both standard ST and PT, samples  of interest correspond to samples obtained at $\tau=1$ i.e.,  $P(\boldsymbol{\theta} \mid \boldsymbol{Y},\tau=1)$.

Marginal likelihood estimates, which are crucial to Baye-sian model comparison and selection, can be obtained using thermodynamic integration (TI) based on the samples obtained in all $\text{T}$ chains of PT \citep{friel2008marginal,calderhead2009estimating}. TI approximates the log marginal likelihood, $\log{P(\boldsymbol{Y})}$, by numerically solving the integral 
$\int\limits_{0}^{1} E_{\boldsymbol{\theta} \mid \tau,\boldsymbol{Y}}[\log(P(\boldsymbol{Y} \mid  \boldsymbol{\theta} ))]d \tau$. As with any numerical integration scheme, accuracy of the integral depends on the discretization, here the number and location of $\tau$ values used in PT. Approximating the thermodynamic integral using PT approximations produces biased marginal likelihood estima-tes \citep{calderhead2009estimating}.  In this paper, we propose a new ST algorithm variant that also solves the Thermodynamic Integral without needing the user to select and tune the number and location of discrete $\tau$ values.

In order to update the inverse temperature parameter, standard ST requires the user to select a prior for $\tau$. Choice of $P(\tau)$ typically requires estimates of the normalizing constant, 
$z(\boldsymbol{Y} \mid \tau)=\int\limits_{\Theta} P(\boldsymbol{Y} \mid \boldsymbol{\theta})^{\tau}P(\boldsymbol{\theta}) d\boldsymbol{\theta}$,  which reduces to a finite dimensional problem if ST is performed over a fixed discretized sequence of $\tau$ values rather than treating $\tau$ as a continuous variable. Consequently, in standard ST, the temperature schedule and the normalizing constants must be estimated through preliminary runs.

Several other methods have been proposed for obtaining normalizing constants \citep{ThompGeyer}: iterative adjustment, Metropolis-Coupled Markov Chain Monte Carlo (MCMCMC) \citep{Geyer}, and stochastic approximation \citep{wasan1969stochastic,wang2001efficient}. Stochastic approximation could
be embedded within standard ST to adaptively estimate the normalizing constants, but this still requires preliminary runs to learn the temperature schedule \citep{atchade2004wang}. Extending this further, the Parallel Adaptive Wang-Landau (PAWL) can be embedded in ST \citep{bornn2013adaptive,bornn2014pawl} to automatically learn the temperature schedule using an adaptive binning strategy. This approach is well suited to parameter estimation but does not resolve the discretization concerns when applied to TI. 

Instead of performing preliminary runs to obtain the most suitable discretized temperature schedule and normalizing constants thereof, we propose a new ST algorithm with a continuous inverse temperature variable defined on $[0,1]$. We remove the requirement for calculating normalizing constants through imposition of the constraint that the distribution of the $\tau$ when profiling over $\boldsymbol{\theta}$ is Uniform. This constraint defines a formula for $P(\tau)$. In the rest of the paper we refer to the new ST algorithm as `Simulated Tempering Without Normalizing Constants' (STWNC).  By sampling across a continuous temperature scale, samples from STWNC can be used for thermodynamic integral estimates of the marginal likelihood.  

In PT and ST, samples from the target distribution are retained whenever $\tau=1$ which is possible (though typically inefficient in standard ST) because of the discretization of the domain of $\tau$.  However, in STWNC the temperature parameter is a continuous variable and therefore $P(\tau=1)=0$. To sample from the target distribution while maintaining a continuously valued $\tau$, we embed STWNC within the PT framework. The resulting PT and STWNC hybrid algorithm named `Parallel Tempering - Simulated Tempering Without Normalizing Constants' (PT-STWNC), runs PT with $\text{T}=2$ chains, each targeting a different goal. The first chain draws samples at continuous temperatures via STWNC. We refer to the first chain as a `tempered' chain. The second chain draws samples from the target distribution where $\tau=1$. We refer to the second chain as a `target' chain in the rest of our presentation.

The  remainder of the paper is organized as follows. Section \ref{sec:PT&ST} gives  an overview of tempering methods with emphasis on PT and standard ST as well as a review of modern variants of stochastic approximation.  Section \ref{sec:STnew} proposes the new STWNC algorithm and provides detailed explanation of how it removes the need for temperature dependent normalizing constants. In Section \ref{sec:PT_STWNC} the hybrid PT-STWNC, which embeds STWNC within a Parallel Tempering algorithm is proposed. Section \ref{sec:TI} presents an overview of marginal likelihood approximation via thermodynamic integration and describes marginal likelihood estimation via STWNC. Sections \ref{sec:Example1} - \ref{sec:SIR} illustrate the PT-STWNC algorithm using several examples: a mixture model of two Gaussian distributions, a mixture model applied to Galaxy velocity data, a Susceptible-Infected-Recovered (SIR) epidemiological ordinary differential Equations (ODE) model. Section \ref{sec:Conclusions} provides discussion.

\section{Tempering Methods}\label{sec:PT&ST}
This section outlines Parallel and Simulated Tempering methods.

\subsection{Parallel tempering}

Parallel Tempering (PT), also known as replica exchange or Population MCMC, is a sampling algorithm designed to improve the dynamic properties of the MCMC samplers especially when it comes to exploring a posterior with many isolated modes. 

Given the likelihood $ P(\mathbf{Y} \mid \boldsymbol{\theta})$,  and prior distribution $P(\theta)$,  the inverse temperature sequence  $0 \leq \tau_{1}<,...,<\tau_{\text{T}}=1 $ defines the $\text{T}$ approximations to the target posterior distribution:
\begin{equation}  P_{t}(\boldsymbol{\theta} \mid \mathbf{Y}) =  \frac{P(\mathbf{Y} \mid \boldsymbol{\theta})^{\tau_{t}} P(\boldsymbol{\theta})}{z(\mathbf{Y} \mid \tau_{t})}, \mbox{ } t\in\{1,..,\text{T}\}, \label{eq:tempered} \end{equation}
where 

\begin{equation}z(\mathbf{Y} \mid \tau_{t})=\int\limits_{\boldsymbol{\Theta}}P( \mathbf{Y} \mid \boldsymbol{\theta} )^{\tau_{t}} P(\boldsymbol{\theta})d\boldsymbol{\theta}\label{eq:norml}\end{equation} 
is the temperature dependent normalizing constant of the $t^{th}$ approximation to the target posterior distribution.
The parameter $\tau$ controls the contribution of the likelihood to the equation (\ref{eq:tempered}), thus enabling the sampler to move  easily and explore the parameter space when $\tau$ is low-valued and to remain further within the basin of attraction of a local mode when $\tau$ is high-valued.

At each iteration PT performs one of two steps: a mutation step where, for example, Metropolis-Hastings (MH) is used to jitter $\boldsymbol{\theta}$ from each of the $\text{T}$ chains independently, and an exchange step where, with some probability two chains $t$ and $l$ are randomly chosen to exchange their parameters $\boldsymbol{\theta}_{t}$ and  $\boldsymbol{\theta}_{l}$. The exchange is accepted with probability:

\small
\begin{eqnarray} 
& &  \min\left\{1, \frac{P_{t}(\boldsymbol{\theta}_{l} \mid \mathbf{Y})} {P_{t}(\boldsymbol{\theta}_{t} \mid \mathbf{Y})}\frac{P_{l}(\boldsymbol{\theta}_{t} \mid \mathbf{Y})}{P_{l}(\boldsymbol{\theta}_{l} \mid \mathbf{Y})} \right\} \nonumber\\
\nonumber\\
& = &    \min\left\{1, \frac{P(\mathbf{Y} \mid \boldsymbol{\theta}_l)^{\tau_{t}} P(\boldsymbol{\theta}_l)P(\boldsymbol{Y} \mid \boldsymbol{\theta}_{t})^{\tau_{l}}P(\boldsymbol{\theta}_t)z(\boldsymbol{Y} \mid \tau_{l})z(\mathbf{Y} \mid \tau_{t})}{z(\mathbf{Y} \mid \tau_{t}) z(\mathbf{Y} \mid \tau_{l})P( \mathbf{Y} \mid \boldsymbol{\theta}_l )^{\tau_{l}} P(\boldsymbol{\theta}_l)P(\mathbf{Y} \mid \boldsymbol{\theta}_t)^{\tau_{t}} P(\boldsymbol{\theta}_t)} \right\} \nonumber \\
\nonumber\\
& =&  \min\left\{1, \frac{P(\mathbf{Y} \mid \boldsymbol{\theta}_l)^{\tau_{t}} P(\mathbf{Y} \mid \boldsymbol{\theta}_t)^{\tau_{l}}  }{P(\mathbf{Y} \mid \boldsymbol{\theta}_l  )^{\tau_{l}} P(\mathbf{Y}  \mid \boldsymbol{\theta}_t )^{\tau_{t}} } \right\}. \nonumber \\ \nonumber 
\end{eqnarray}
\normalsize

Note the cancellation of normalizing constants $z(\mathbf{Y} \mid \tau_{l})$ and $z(\mathbf{Y} \mid \tau_{t})$ when swapping parameters between two different temperature based chains. The cancellation of normalizing constants and the ease of movement between modes at low values of $\tau$ has made PT into a widely used algorithm. Samples from the $\text{T}^{th}$ chain correspond to samples from the target distribution. 

\subsection{Simulated tempering} \label{sec:ST}

Simulated tempering (ST), also known as serial tempering, is a single chain sampling method where the posterior parameter space is augmented by including the temperature, $\tau$, as a random variable. Consequently ST requires specification of the prior $P(\tau)$. As with PT, $\tau$ controls the influence of the likelihood on $P(\boldsymbol{\theta} \mid \mathbf{Y},\tau)$.   However, in ST different temperatures are explored in a  random walk through the joint distribution of $\boldsymbol{\theta}$ and $\tau$ :   
\begin{equation}  P(\boldsymbol{\theta}, \tau \mid \mathbf{Y}) =\frac{ P(\mathbf{Y} \mid \boldsymbol{\theta})^{\tau} P(\boldsymbol{\theta})P(\tau)}{P(\mathbf{Y})},\label{eq:STposterior}  \end{equation}
where the normalizing constant for the joint density is:

\[P(\mathbf{Y})=\int\limits_0^1\int\limits_{\boldsymbol{\Theta}} P(  \boldsymbol{Y} \mid \boldsymbol{\theta})^{\tau} P(\boldsymbol{\theta})P(\tau)d\boldsymbol{\theta} d\tau.\] 

Similar to PT, samples that arise at $\tau=1$ are samples from target distribution, however, when $\tau$ is continuous, $P(\tau=1)=0$.

\subsubsection{Prior for $\tau$} \label{sec:ChoosePrior}
In order for standard ST to mix well, a carefully chosen prior $P(\tau)$ needs to be defined.
Following  \cite{ThompGeyer}, the prior for the inverse temperature can be found by examining the marginal distribution of $\tau$: 

\begin{equation}\begin{array}{ll}        
P(\tau \mid \boldsymbol{Y}) =\int\limits_{\boldsymbol{\Theta}} P(\tau,\boldsymbol{\theta} \mid \boldsymbol{Y}  ) d\boldsymbol{\theta}   
&\propto P(\tau)  \int\limits_{\boldsymbol{\Theta}}  P(\boldsymbol{Y} \mid \boldsymbol{\theta}  )^{\tau}  P(\boldsymbol{\theta})  d\boldsymbol{\theta}
\\ 
\\ &\propto  P(\tau)z(\boldsymbol{Y} \mid \tau) \\   \label{eq:margTau}
\end{array}\end{equation}
where  
$z(\boldsymbol{Y} \mid \tau)=\int\limits_{\boldsymbol{\Theta}} P(\boldsymbol{Y} \mid \boldsymbol{\theta}  )^{\tau}  P(\boldsymbol{\theta})  d\boldsymbol{\theta}$ is the $\tau$ dependent normalizing constant of the conditional posterior distribution $P(\boldsymbol{\theta} \mid \tau, \boldsymbol{Y})$. If the prior for $\tau$ is chosen to be approximately proportional to the inverse normalizing constant, that is, if
\begin{eqnarray}        
P(\tau) \propto \frac{1}{z(\boldsymbol{Y} \mid \tau)}, \label{eq:stSTpriortau}
\end{eqnarray}
then the marginal distribution of $\tau \mid \boldsymbol{Y} \sim \mbox{Uniform(0,1)}$. 
Following  \cite{ThompGeyer}, the standard ST algorithm 
starts with $n$ tempered distributions at $\tau_{i}=\frac{i}{T}$ where $i=0,..,n \leq T$, and iterates between adjusting the prior of $\tau$ and adjusting the inverse temperature spacing until a desired rate for transitions between the interpolating distributions is met. Afterwards, new inverse temperatures are added, and a new iterative cycle of adjusting prior and inverse temperature spacing is started for the newly added temperature set.

\subsection{Outline of the Standard Simulated Tempering}

A Markov Chain using standard ST updates parameters through Gibbs steps, updating $\boldsymbol{\theta} \mid\tau, \boldsymbol{Y}$ and $\tau\mid \boldsymbol{\theta}, \boldsymbol{Y}$ in turn.  In the first kind of transition, a fixed temperature mutation step is typically a Metropolis Hastings step identical to what would be used to sample one of the fixed temperature chains $P_{t}(\boldsymbol{\theta} \mid \boldsymbol{Y})$ in PT.

Updating  $\tau\mid \boldsymbol{\theta}^{(i+1)}, \boldsymbol{Y}$ through Metropolis Hastings occurs by proposing a value $\tau^*$ sampled from a symmetric distribution (for expositional simplicity). Using the prior in (\ref{eq:stSTpriortau}), the value $\tau^*$ is accepted (setting $\tau^{(i+1)}=\tau^*)$ with probability:

\begin{eqnarray}
& & \min\left\{1,\frac{  P(\boldsymbol{\theta}^{(i+1)},\tau^{*} \mid \boldsymbol{Y})  }{   P(\boldsymbol{\theta}^{(i+1)},\tau^{(i)} \mid  \boldsymbol{Y}) }    \right\}  \nonumber \\
\nonumber \\       & = & \min\left\{1,\frac{ P(\boldsymbol{Y} \mid \boldsymbol{\theta}^{(i+1)}  )^{\tau^*} P(\boldsymbol{\theta}^{(i+1)})P(\tau^*)P(\boldsymbol{Y})}{P( \boldsymbol{Y} \mid \boldsymbol{\theta}^{(i+1)} )^{\tau^{(i)}} P(\boldsymbol{\theta}^{(i+1)})P(\tau^{(i)})P(\boldsymbol{Y})}    \right\} \nonumber \\
\nonumber \\ 
\nonumber \\  &= &\min\left\{1,\frac{ P(\boldsymbol{Y} \mid \boldsymbol{\theta}^{(i+1)}  )^{\tau^*} z(\boldsymbol{Y} \mid \tau^{(i)}) }{P( \boldsymbol{Y} \mid \boldsymbol{\theta}^{(i+1)})^{\tau^{(i)}}z(\boldsymbol{Y} \mid \tau^*)}  \right\}.    \label{eq:STaccept} \end{eqnarray}

Consequently, the acceptance probability relies on temperature dependent normalizing constant $z(\boldsymbol{Y} \mid \tau)$.  
Normalizing constants, also referred to as weights of the sequence of distributions \citep{neal1996sampling}, are generally unknown and finding suitable values requires pilot runs -- defeating the purpose of using standard ST. Information from pilot runs also cannot be recycled when additional data becomes available because $P(\tau)$ depends on $\boldsymbol{Y}$. To make the normalizing constant problem simpler, $\tau$ is discretized instead of being continuously valued.

\subsection{Related algorithms}	  \label{sec:RelatedAlg}

The Wang-Landau (WL) algorithm \citep{wang2001efficient}, a stochastic approximation algorithm, could be used within standard ST to  automatically obtain $P(\tau)$       \citep{ThompGeyer,atchade2004wang}. Standard WL algorithm requires a predefined temperature schedule $ 0 \leq \tau_{1}<,..,<\tau_{\text{T}} =1 $  corresponding to partitioning the temperature state space $\mathbb{T}$ into $\text{T}$ different regions i.e., bins $\text{E}_{1},..,\text{E}_{\text{T}}$. The goal is to construct a chain that could spend the same time in each $\text{E}_{t}$. The moves inside the $\text{E}_{t}$ are performed with a standard Metropolis-Hastings (MH) algorithm with target distribution $\pi$. The WL algorithm recursively re-weights $\pi$ in $\text{E}_{t}$ by a factor $\phi_{i}(t)$. Hence, given $\tau^{(i)}$ and some unnormalized weights $\phi_{i}$, the $\tau^{(i+1)}$ can be sampled using a MH algorithm with invariant density proportional to $\sum\limits_{t=1}^{\text{T}}\frac{\pi(\tau)}{\phi_{i}(t)}\mathbb{I}_{\text{E}_{t}}(\tau)$, where $\mathbb{I}$ is the indicator function. The normalizing constants at the $i$-th iteration $\phi_{i}(t)= z(\textbf{Y} \mid \tau_{t}^{(i)})$ are updated for each of the bins $t \in \{1,..,\text{T}\}$ until a predefined criterion is met. This criterion, also known as 'flat histogram', ensures that proportions of visits of the chain in each of the bins $\text{E}_{1},..,\text{E}_{\text{T}}$ are approximately equal to $\text{T}^{-1}$. The updating rule for the normalizing constants yields $\phi_{i+1}(t)= \phi_{i}(t)(1+\gamma_{i}\mathbb{I}_{\text{E}_{t}}(\tau^{(i+1)}))$, 
where $\gamma_{i}$ is a learning rate which decreases stochastically until the criterion of 'flat histogram' of visit frequency to the bins $\text{E}_{1},..,\text{E}_{\text{T}}$ is met.

The Parallel Adaptive WL (PAWL) algorithm  \citep{bornn2013adaptive} removes the need for preliminary runs of WL to learn the optimal partitioning of the state space by exploiting an  adaptive binning strategy. The adaptive binning strategy requires initial bins  and  bin range to be specified by the user. Since $\phi_{i}(t)$ represent normalizing constant for the $t$-th bin, the adaptive binning strategy maintains uniformity within a bin to allow within-bin movement. This is achieved by determining presence of heavy tails in the distribution of the samples within each bin. If the distribution is skewed towards the left side, then the sampler will have difficulty moving to the neighboring bin on the left. 
Hence, the binning strategy divides the bin into two chains by the middle point of the bin, and then it measures discrepancy between the chains using a ratio of the number of points in any of the two chains and the number of points within the bin. If this ratio is close to $50 \%$ then the histogram of the within-bin distribution is close to uniform. Hence, if the ratio is below some threshold, for example $25 \%$, two new bins are created using the middle point of the former bin, and otherwise, the bin remains unchanged. One can specify  the threshold to be $50 \%$, but then the number of newly created bins will be larger. PAWL checks if the bins have to be split until the 'flat histogram' criterion is met, and afterwards the bin splitting stops since the sampler can move easily between the bins. Embedding PAWL within standard ST has been explored  with a purpose to automate the two input requirements for ST: choice of temperature schedule and calculation of $P(\tau)$         \citep{bornn2014pawl}.

The Equi-Energy (EE) sampler \citep{Kou2006}, which utilizes temperature-energy duality, also allows wide moves by performing jumps between the states with similar energy levels.    The EE, which is a powerful sampling and estimation methodology  that addresses multi-modality in high-dimensional target distributions, provides estimates of expectations under any fixed temperature. However, the discrete nature of the temperature in EE sampler requires careful tuning of the temperature schedule in order for the EE samples to be applicable to thermodynamic integration.

\section{Simulated Tempering Without Normalizing Constants } \label{sec:STnew}

The standard ST is not widely used in practice because of the challenges that arise from finding suitable values of the unknown normalizing constants. The proposed Simulated Tempering Without Normalizing Constants (STWNC) algorithm removes the dependence on normalizing constants in the acceptance ratio in (\ref{eq:STaccept}) by the way we define the prior for $\tau$.
As with ST, STWNC is still moving through two kinds of transitions: updating $(\boldsymbol{\theta}\mid\tau, \boldsymbol{Y})$; and updating $(\tau\mid \boldsymbol{\theta}, \boldsymbol{Y})$. The nuisance parameter $\tau$ is not of inferential interest, consequently its prior $P(\tau)$ can be selected for algorithmic convenience.  Our algorithm therefore chooses $P(\tau$) by imposing a constraint that the profile posterior distribution of $\tau$ while maintaining $\boldsymbol{\theta} \mid\tau,\boldsymbol{Y}$ at its maximum value results in a  uniform distribution. Using this constraint we derive a formula for the prior of $\tau$ which is computationally inexpensive and does not require preliminary runs of the algorithm or discretization of $\tau$. In the remainder of this section we describe the derivation of formula for the prior of $\tau$.

We impose the constraint that the posterior distribution of $\tau$ while profiling over $\boldsymbol{\theta}$ is uniform: 

\begin{equation}
\boldsymbol{\theta}_{\max}(\tau)=\arg\max_{\boldsymbol{\theta}}{P(\boldsymbol{\theta} \mid \tau, \boldsymbol{Y} )}, \label{EQ:optim}
\end{equation}

\begin{equation}
P(\tau, \boldsymbol{\theta}_{\max}(\tau) \mid \boldsymbol{Y})  =U(0,1). \label{EQ:profUnif}\end{equation}
Equation (\ref{EQ:profUnif})  implies that the joint posterior distribution of $\tau$ and ${\theta}_{\max}(\tau)$, $P(\tau,  \boldsymbol{\theta}_{\max}(\tau) \mid \boldsymbol{Y})$ has a ridge of constant maximum height from $\tau=0$ to $\tau=1$. 
In other words for any $\tau_i,\tau_j\in[0,1]$,

\begin{equation}
P(\tau_{i},  \boldsymbol{\theta}_{\max}(\tau_{i}) \mid \boldsymbol{Y})= P(\tau_{j},  \boldsymbol{\theta}_{\max}(\tau_{j}) \mid \boldsymbol{Y}).\label{EQ:cprof}\end{equation}

To derive the prior of $\tau$, we expand the profile posterior distribution given by the equation (\ref{EQ:profUnif}):

\small
\begin{eqnarray} 
P( \tau,\boldsymbol{\theta}_{\max}(\tau)  \mid \boldsymbol{Y}) & = & \nonumber \\
  \frac{P(\boldsymbol{Y} \mid \boldsymbol{\theta}_{\max}(\tau)  )^{\tau}P(\boldsymbol{\theta}_{\max}(\tau))P(\tau)}{P_{\text{prof}}(\boldsymbol{Y})} \nonumber \\ \label{EQ:joint}
\end{eqnarray} 
\normalsize
where $P(\boldsymbol{Y} \mid \boldsymbol{\theta}_{\max}(\tau))^{\tau}$ is tempered profile likelihood,
$P_{\text{prof}}(\boldsymbol{Y}) = \int\limits_{0}^{1} P(\boldsymbol{Y} \mid  \boldsymbol{\theta}_{\max}(\tau) )^{\tau}P(\boldsymbol{\theta}_{\max}(\tau))P(\tau) d\tau$ is the normalizing constant of the profile posterior distribution of $\tau$ over $\boldsymbol{\theta}$, $P(\tau)$ is the prior of the inverse temperature and $P(\boldsymbol{\theta}_{\max}(\tau))$ is prior of $\boldsymbol{\theta}$ evaluated at $\boldsymbol{\theta}_{\max}(\tau)$.
Expressing $P(\tau)$  by rearranging (\ref{EQ:joint}) gives

\begin{eqnarray}
P(\tau)  =  \frac{P( \tau, \boldsymbol{\theta}_{\max}(\tau)  \mid \boldsymbol{Y})P_{\text{prof}}(\boldsymbol{Y})}{P(\boldsymbol{Y} \mid  \boldsymbol{\theta}_{\max}(\tau) )^{\tau}P(\boldsymbol{\theta}_{\max}(\tau)   ) }.     \label{EQ:ptau}
\end{eqnarray}

As a direct consequence of the constraint that profile posterior distribution of $\tau$ while maintaining $\boldsymbol{\theta} \mid \tau, \boldsymbol{Y}$ at its maximum value is uniform, i.e., (\ref{EQ:profUnif}) and (\ref{EQ:cprof}), the numerator in the formula for prior of $\tau$ in (\ref{EQ:ptau}) is constant with respect to $\tau$.

Using the equation (\ref{EQ:ptau}) as prior for $\tau$, the STWNC acceptance ratio for a proposed $\tau^*$ alters (\ref{eq:STaccept}) into:

\small
\begin{eqnarray} 
\nonumber
& &  \min\left\{1,\frac{ P( \boldsymbol{Y} \mid \boldsymbol{\theta} )^{\tau^*} P(\tau^*)}{P(\boldsymbol{Y} \mid \boldsymbol{\theta} )^{\tau^{(i)}} P(\tau^{(i)})}    \right\} \\ \nonumber  \\ \nonumber \\ \nonumber
&=& \min  \left\{ 1,\frac{P( \boldsymbol{Y}  \mid \boldsymbol{\theta})^{\tau^*} 
P( \tau^{*}, \boldsymbol{\theta}_{\max}(\tau^{*})  \mid \boldsymbol{Y})
P_{\text{prof}}(\boldsymbol{Y})}{  P( \boldsymbol{Y}  \mid  \boldsymbol{\theta}_{\max}(\tau^*))^{\tau^*}P(\boldsymbol{\theta}_{\max}(\tau^*)   ) }    \right. \times \nonumber \\ \nonumber \\   
&&  \qquad \left.\frac{   P(\boldsymbol{Y} \mid  \boldsymbol{\theta}_{\max}(\tau^{(i)}) )^{\tau^{(i)}}P(\boldsymbol{\theta}_{\max}(\tau^{(i)})   )  }  {P(\boldsymbol{Y}  \mid \boldsymbol{\theta} )^{\tau^{(i)}} 
P( \tau^{(i)}, \boldsymbol{\theta}_{\max}(\tau^{(i)})  \mid \boldsymbol{Y})P_{\text{prof}}(\boldsymbol{Y})}  \right\}\nonumber \\ \nonumber
\\ \nonumber \\  \nonumber
&=& \min\left\{1,\frac{P(\boldsymbol{Y} \mid \boldsymbol{\theta})^{\tau^{*}}  P(\boldsymbol{Y} \mid \boldsymbol{\theta}_{\max}(\tau^{(i)})  )^{\tau^{(i)}}P(\boldsymbol{\theta}_{\max}(\tau^{(i)}))}{P(\boldsymbol{Y} \mid \boldsymbol{\theta}  )^{\tau^{(i)}}P( \boldsymbol{Y} \mid  \boldsymbol{\theta}_{\max}(\tau^{*}))^{\tau^{*}}P(\boldsymbol{\theta}_{\max}(\tau^{*}))}  \right\}. 
\\ \label{eq:Updtau}
\end{eqnarray}
\normalsize

The acceptance ratio in equation (\ref{eq:Updtau}) does not depend on the temperature dependent normalizing constant $z(\boldsymbol{Y} \mid \tau)$
thus eliminating the need for discrete temperatures, normalizing constant estimates, and tuning of bin widths.

The formula for the prior of $\tau$ imposes the property that the maximal contour of the joint posterior distribution is continuous for $\tau \in [0,1]$.  However, calculation of  $\boldsymbol{\theta}_{\max}(\tau^{*}) $ requires optimization for each proposed value of $\tau^*$.  While this initially may be prohibitive, the optimization cost decays with increasing iterations because of the improved starting points available for the optimizer.  At iteration $i$, with proposed $\tau^*$ we initialize the optimizer at the point $\boldsymbol{\theta}_{\max}(\tau^{(k)})$ for the $k<i$ that minimizes: $||\tau^{(k)}-\tau^*||$.  As $i\rightarrow\infty$ the optimization is initialized arbitrarily close to the optimum.  The computational cost of evaluating the prior for $\tau$ is reduced, but some computational overhead remains by instead searching through previously obtained values for initialization.  This approach was implemented in the examples in this paper and is incorporated into the Pseudo-code of STWNC is given in Algorithm \ref{alg:STWNC}.

For long running Markov Chains, further computational reduction may be possible within the MCMC iterations by approximating $P(\tau)$ with an iteratively updated Gaussian process approximation \citep{ConradMarzoukPillaiSmith}, or by approximating the manifold of $\theta_{\max}(\tau)$ based on offline evaluations thereof.  In this latter approach, before beginning MCMC iterations, optimization of $\theta(\tau)$ is performed over a grid of values of $\tau$.  An interpolator is defined so as to replace the optimization of $\theta(\tau)$ within each MCMC iteration with evaluation of the interpolator instead.  After this initial computational cost, the time per iteration is equivalent to that of parallel tempering with the same number of parallel chains.  This approach is explored as proof of concept and not in a general sense because the precision of the approximation depends on the quality of the interpolator which in turn will be impacted by smoothness and continuity of the manifold, properties which are difficult to characterize or guarantee given the multi-modal nature of the likelihoods in this manuscript.  This latter strategy is demonstrated as proof of concept in Section \ref{sec:accel}.


\begin{algorithm}[]   
	\caption{Simulated Tempering Without Normalizing Constants (STWNC)}
	\textbf{Goal:}  Update $\boldsymbol{\theta}$ and $\tau$ from $P(\boldsymbol{\theta},\tau\mid \boldsymbol{Y})$, where $\tau \in [0,1]$ is continuous.
	\\
	Initialize the algorithm with $i=0$ and some values for ($\boldsymbol{\theta}^{(i)}$, $\tau^{(i)}$); define $\text{N}$ - the number of iterations.
	\begin{algorithmic}
		\For{$i = 1 : \text{N}$}
		\State \textbf{ Transition 1:}  update ($\boldsymbol{\theta} \mid \tau^{(i)}$);  
		\begin{enumerate} 
			\item[] propose a ($\boldsymbol{\theta}^{*}$); 
			\item[] calculate the MH ratio $\alpha_{\boldsymbol{\theta}}$ and accept or reject $\boldsymbol{\theta}^{*}$ :
			\begin{eqnarray}\alpha_{\boldsymbol{\theta}} = \frac{P(\boldsymbol{\theta}^*,\tau^{(i)}\mid \boldsymbol{Y})}{P(\boldsymbol{\theta}^{(i)},\tau^{(i)}\mid \boldsymbol{Y})}  
			=\frac{P(\boldsymbol{\theta}^* \mid \tau^{(i)}, \boldsymbol{Y})}{P(\boldsymbol{\theta}^{(i)} \mid \tau^{(i)}, \boldsymbol{Y})}; \nonumber \label{alpha1}
			\end{eqnarray}
			\item[]  sample a $u_{\boldsymbol{\theta}}\sim U(0,1)$;
		\end{enumerate}
		\If {$u_{\theta}<\alpha_{\boldsymbol{\theta}}$}  set $(\boldsymbol{\theta}^{(i+1)},\tau^{(i)} ) \leftarrow (\boldsymbol{\theta}^* ,\tau^{(i)} )$,  \Else{  retain $(\boldsymbol{\theta}^{(i+1)} ,\tau^{(i)} ) \leftarrow (\boldsymbol{\theta}^{(i)} ,\tau^{(i)} ) $}; \EndIf
		\State \textbf{Transition 2:}  update ($\tau \mid \boldsymbol{\theta}^{(i+1)}$); 
		\begin{enumerate}
			\item[] propose a $\tau^*$; 
			\item[]  find the $k<i$ that minimizes: $||\tau^{(k)}-\tau^*||$
                         \item[]  optimize $\boldsymbol{\theta}_{\max}(\tau^{*}) $ by initializing from $\boldsymbol{\theta}_{\max}(\tau^{(k)}) $
                         \item[] calculate the MH ratio $(\alpha_{\tau})$ and accept or reject   $(\boldsymbol{\theta}^{(i+1)},\tau^*)$:
			\begin{eqnarray} 
			\small
			\alpha_{\tau} &=& \frac{P(\boldsymbol{\theta}^{(i+1)},\tau^*\mid \boldsymbol{Y})}{P(\boldsymbol{\theta}^{(i+1)},\tau^{(i)} \mid \boldsymbol{Y})}  \nonumber \\
			& = & \frac{P(\boldsymbol{Y} \mid  \boldsymbol{\theta} )^{\tau^*}  P(\boldsymbol{Y} \mid  \boldsymbol{\theta}_{\max}(\tau^{(i)})  )^{\tau^{(i)}}P(\boldsymbol{\theta}_{\max}(\tau^{(i)})   )}{ P(\boldsymbol{Y} \mid \boldsymbol{\theta}  )^{\tau^{(i)}}P( \boldsymbol{Y} \mid \boldsymbol{\theta}_{\max}(\tau^*) )^{\tau^{*}}P(\boldsymbol{\theta}_{\max}(\tau^*)   ) } \nonumber
			\label{alpha2}   \end{eqnarray} 
			\item[]   sample a $u_{\tau}\sim U(0,1)$;
		\end{enumerate}
		\If {$u_{\tau}<\alpha_{\tau}$}  set $(\boldsymbol{\theta}^{(i+1)},\tau^{(i+1)}) \leftarrow (\boldsymbol{\theta}^{(i+1)},\tau^*)$,  \Else { retain $(\boldsymbol{\theta}^{(i+1)},\tau^{(i+1)}) \leftarrow (\boldsymbol{\theta}^{(i+1)},\tau^{(i)})$}; \EndIf
		\EndFor
		\State  \textbf{Return}: a single chain of samples $\{\boldsymbol{\theta},\tau\}$.
	\end{algorithmic}
	\label{alg:STWNC}
\end{algorithm}

\section{Parallel Tempering via Simulated Tempering without Normalizing Constants  algorithm}  \label{sec:PT_STWNC}

In Parallel Tempering via Simulated Tempering without Normalizing Constants (PT-STWNC), the inverse temperature is a continuous parameter and $P(\boldsymbol{\theta} \mid \boldsymbol{Y})=\int\limits_{0}^{1}  P(\boldsymbol{\theta}, \tau \mid \boldsymbol{Y})d\tau$ is a marginal distribution, which does not coincide with the target distribution.  Instead, the target distribution is  $P(\boldsymbol{\theta} \mid \boldsymbol{Y}, \tau=1)$.  To  obtain samples from the target distribution we run PT with $T=2$ defined as follows:
\begin{itemize}
	\item  The first PT chain, called the tempered chain, updates $\boldsymbol{\theta}$ and $\tau$ via STWNC using $P(\boldsymbol{\theta},\tau\mid \boldsymbol{Y})$ 
	\item  The second PT chain, called the target chain, updates $\boldsymbol{\theta}$ via standard Metropolis-Hastings or Gibbs sampling  from the target distribution $P(\boldsymbol{\theta}\mid \boldsymbol{Y},\tau=1)$. 
\end{itemize}

As in standard PT, the PT-STWNC algorithm goes through two transitions: exchange and mutation. In the exchange step, the exchange between the two chains is proposed. If the exchange was accepted, then the two chains swap the parameter values between each other, and otherwise the two chains go through mutation steps. In the mutation step, the first chain updates the parameters of interest at different temperatures via STWNC; the second chain updates the parameters of interest at $\tau=1$ via Metropolis-Hastings. 
Pseudo-code of  PT-STWNC is given in Algorithm \ref{alg:PT-STWNC}.

\begin{algorithm}[]
	\caption{Parallel Tempering using Simulated Tempering Without Normalizing Constants (PT-STWNC)}
	Initialize two parallel chains: 
	`tempered' chain -- initialize the algorithm with some values for ($\boldsymbol{\theta}_{1}^{(i)}$, $\tau_{1}^{(i)}$) for $i=0$ and
	`target' chain -- initialize the algorithm with values for ($\boldsymbol{\theta}_{2}^{(i)}, \tau_{2}=1$) for $i=0$; $N$ - the number of iterations.
	
	\begin{algorithmic}
		\For{$i = 1:N$}
		\State with probability $\rho$, propose an exchange between the two chains;
		
		\If {exchange is proposed}
		\State calculate the exchange probability: 
		\begin{equation}
		\alpha_{\phi}=\frac{P( \boldsymbol{Y} \mid  \boldsymbol{\theta}^{(i)}_{2}   )^{\tau_{1}^{(i)}}P(\boldsymbol{Y} \mid \boldsymbol{\theta}^{(i)}_{1}  )^{\tau_{2}=1} } {P(\boldsymbol{Y} \mid \boldsymbol{\theta}^{(i)}_{1})^{\tau_{1}^{(i)}}P(\boldsymbol{Y} \mid \boldsymbol{\theta}^{(i)}_{2} )^ {\tau_{2}=1} }
		\end{equation}
		\State   sample a $u\sim U(0,1)$;
		
		\If {$u<\alpha_{\phi}$}  
		
		\State   swap  states of the two chains:
		\State   $(\boldsymbol{\theta}_{1}^{(i+1)}, \boldsymbol{\theta}_{2}^{(i+1)}) \leftarrow (\boldsymbol{\theta}_{2}^{(i)}, \boldsymbol{\theta}_{1}^{(i)})$ ;
		\Else
		\State  retain $(\boldsymbol{\theta_{1}}^{(i+1)},\boldsymbol{\theta}_{2}^{(i+1)}) \leftarrow (\boldsymbol{\theta}_{1}^{(i)},\boldsymbol{\theta}_{2}^{(i)})$;
		\EndIf
		\Else {\ \ mutation is performed:}
		\State Update the `tempered' chain, i.e., update  ($\boldsymbol{\theta}_{1}^{(i+1)}$,$\tau_{1}$) via STWNC using Algorithm \ref{alg:STWNC};
		\State Update the `target' chain, i.e., update ($\boldsymbol{\theta}_{2}^{(i+1)}$, $\tau_{2}=1$) via Metropolis-Hastings;
		\EndIf
		\EndFor
		\State  \textbf{Return}:  samples from the 'tempered' chain  $\{\boldsymbol{\theta}_{1},\tau_{1}\}$, and samples from the 'target' chain 
		$\{\boldsymbol{\theta}_{2},\tau_{2}=1\}$.
	\end{algorithmic}
	\label{alg:PT-STWNC}
\end{algorithm}

\section{Estimation of Marginal Likelihoods via thermodynamic integration}  \label{sec:TI}

Posterior model probabilities $P(M \mid \boldsymbol{Y})$ provide an intuitive framework for evaluating model $M$ within a model class.  Expanding the model class requires rescaling all posterior model probabilities.  Consequently, comparing models $M_{1}$  and $M_{2}$ is typically performed through the posterior odds,   

\begin{equation}
\frac{P(\text{M}_{1} \mid \boldsymbol{Y})}{P(\text{M}_{2} \mid \boldsymbol{Y})}=\frac{P(  \boldsymbol{Y} \mid \text{M}_{1}  )}{P(  \boldsymbol{Y} \mid \text{M}_{2} )} \frac{P(\text{M}_{1})}{P(\text{M}_{2})}. \label{eq:BF}
\end{equation}
The  ratio of posterior and prior odds,
\begin{equation}
\text{B}_{12}=\frac{P(  \boldsymbol{Y} \mid \text{M}_{1}  )}{P(  \boldsymbol{Y} \mid \text{M}_{2} )}  \label{eq:BF_1} 
\end{equation}
is Bayes Factor of $\text{M}_{1}$ against $\text{M}_{2}$             \citep{kass1995bayes}.

When there are no prior preferences for models, $\text{B}_{12}$ is equal to the posterior odds. The marginal likelihoods for models $\text{M}_{j},j=1,2$,  in  (\ref{eq:BF_1}), are obtained by integrating over the parameter space,

\begin{equation}
P(  \boldsymbol{Y}  \mid \text{M}_{j} )=\int\limits_{\boldsymbol{\Theta}_{j}} P(\boldsymbol{Y}  \mid \boldsymbol{\theta}_{j}, \text{M}_{j})P(\boldsymbol{\theta}_{j} \mid \text{M}_{j} )d\boldsymbol{\theta}_{j},       \label{eq:ordmllik}
\end{equation}
where $\boldsymbol{\theta}_{j}$ are the parameters corresponding to the $j$-th model.  For expositional simplicity we will assume no prior preference for models throughout this paper.  Computing meaningful Bayes Factors requires accurate estimates of the marginal likelihood in (\ref{eq:ordmllik}).  

The Posterior Harmonic Mean estimator (PHM), uses importance sampling to integrate (\ref{eq:ordmllik}) \citep{newton1994approximate, raftery2006estimating} resulting in unbiased marginal likelihood estimates but potentially infinite variance. Steppingstone sampling (SS) \citep{xie2011improving}, which  uses importance sampling to estimate each ratio of normalizing constants of the sequence of interpolating distributions between prior and target distribution, provides reliable marginal likelihood estimates. Alternatively, thermodynamic integration (TI) \citep{friel2008marginal} builds on ideas from path sampling \citep{Gelman1998} to estimate the marginal likelihood via,
\begin{equation}
\log(P(\boldsymbol{Y}\mid \text{M}_{j}))=\int\limits_{0}^{1} \E_{\boldsymbol{\theta} \mid \boldsymbol{Y},\tau,\text{M}_{j}}[\log(P(  \boldsymbol{Y} \mid \boldsymbol{\theta}, \text{M}_{j}))] d\tau,    \label{eq:marglik}
\end{equation}
where the expectation in the integrand is with respect to the tempered posterior distribution in  (\ref{eq:tempered})   \citep{friel2008marginal,calderhead2009estimating}.	
A numerical approximation  to the thermodynamic integral in (\ref{eq:marglik}) is possible through discretization of $\tau$.  In \cite{friel2008marginal}, samples from the discretized tempered posteriors were used from parallel chains  in PT.  At each discretized value of $\tau$,  

$\E_{\boldsymbol{\theta} \mid \boldsymbol{Y},\tau,\text{M}_{j}}[\log(P(\boldsymbol{Y} \mid \boldsymbol{\theta},\text{M}_{j}))]$ is evaluated and the marginal likelihood is approximated by applying a trapezoid rule to numerically integrate over $\tau$,	
\begin{eqnarray}
\log (P(\boldsymbol{Y} \mid \text{M}_{j})) = & \int\limits_{0}^{1} \E_{\boldsymbol{\theta} \mid \boldsymbol{Y},\tau, \text{M}_{j}}[\log(P(  \boldsymbol{Y} \mid \boldsymbol{\theta},\text{M}_{j}))] d\tau   \nonumber 
\\ \approx& \frac{1}{2}\sum\limits_{t=2}^{\text{T}}\Delta\tau_{t}(\E_{t,{\text{M}_{j}}}+\E_{t-1,\text{M}_{j}}),  \label{eq:discrMargPF}
\end{eqnarray}
where $\E_{t,\text{M}_{j}}=\E_{\boldsymbol{\theta} \mid \boldsymbol{Y},\tau_{t},{\text{M}_{j}}}[\log(P( \boldsymbol{Y} \mid \boldsymbol{\theta},\text{M}_{j} ))]$ and $\Delta\tau_{t}= \tau_{t}-\tau_{t-1}$.
The discretized trapezoidal rule in  (\ref{eq:discrMargPF}) was improved by \cite{calderhead2009estimating}   by correcting for integration bias in terms of Kullberg-Leibler divergence, $\text{KL}(p_{t-1,{\text{M}_{j}}} \|  p_{t,{\text{M}_{j}}})$, of $p_{t,{\text{M}_{j}}}$ from  $p_{t-1,{\text{M}_{j}}}$ for model $\text{M}_{j}$,

\begin{eqnarray}
&& \log(P(\boldsymbol{Y} \mid \text{M}_{j}))\approx \underbrace{\frac{1}{2}\sum_{t=2}^{\text{T}}\Delta\tau_{t}(\E_{t,{\text{M}_{j}}}+\E_{t-1,{\text{M}_{j}}})}_{\text{Approximation}}+ \nonumber \\  && \underbrace{\frac{1}{2} \sum_{t=2}^{\text{T}}[\text{KL}(p_{t-1,{\text{M}_{j}}} \|  p_{t,{\text{M}_{j}}})  - \text{KL}(p_{t,{\text{M}_{j}}} \| p_{t-1,{\text{M}_{j}}})]}_{\text{Bias}}, \nonumber \\
\label{eq:discrMarg}
\end{eqnarray}
\normalsize
where $p_{t,{\text{M}_{j}}}$ is the tempered posterior distribution for the model $\text{M}_{j}$ given by the equation (\ref{eq:tempered}). 

The thermodynamic integration via PT applies a numerical integration approximation to Monte Carlo approximations. The approximation error should decrease with number of chains and number of samples in each chain.
As with any numerical integration, discretization over $\tau$ determines the accuracy of the result. Determining the optimal temperature schedule requires preliminary experimentation which contributes to unpopularity of the thermodynamic integration in practice. Based on the idea from path sampling, \citep{Gelman1998}, \cite{calderhead2009estimating} proposed that the temperature schedule could be chosen such that the Monte Carlo variance of the marginal likelihood estimates is minimized. The authors' numerical simulations suggest that the optimal temperature schedule is $\tau_{i}=(\frac{i}{\text{T}})^{5}$ in their situations, which puts more emphasis on values closer to the prior.

\subsection{Computing the marginal likelihood via STWNC }  \label{seq:margPTSTWNC}

Following \cite{friel2008marginal}, in standard ST, samples $\{  (\boldsymbol{\theta}^{1}, \tau_{1}),..,(\boldsymbol{\theta}^{n}, \tau_{n}) \}$  drawn from $P(\boldsymbol{\theta},\tau \mid \boldsymbol{Y},\text{M}_{j})$ can be used to estimate the marginal likelihood by first obtaining  Monte Carlo approximation  $\E_{\boldsymbol{\theta} \mid \boldsymbol{Y}, \tau, \text{M}_{j}}[\log{P(\boldsymbol{Y} \mid  \boldsymbol{\theta}, \text{M}_{j} )}]$, and then solving the thermodynamic integral in  (\ref{eq:discrMargPF}) 
via quadrature.  This is based on the assumption that the prior of $\tau$ is proportional to the temperature-dependent normalizing constant, $P(\tau) \propto z(\boldsymbol{Y} \mid \tau, \text{M}_{j})$. According to \cite{friel2008marginal}, in single chain methods such as ST, the normalizing constant $z(\boldsymbol{Y} \mid \tau, \text{M}_{j})$ varies by orders of magnitude with $\tau$ which leads to poor estimation of the log untempered likelihood $\log{P(\boldsymbol{Y} \mid \boldsymbol{\theta}, \text{M}_{j})}$. Thus in standard ST, small values of $\tau$ do not tend to be sampled with high frequencies.  However, in STWNC, where $\tau$ is continuous, the marginal distribution of $\tau$ tends to spend a lot of time at near zero values 
The marginal distribution of $\tau$ in STWNC coincides with that of the recommended geometric temperature schedule for thermodynamic integration via PT suggested by \cite{calderhead2009estimating}. 

Samples  $\{  (\boldsymbol{\theta}^{1}, \tau_{1}),..,(\boldsymbol{\theta}^{n}, \tau_{n}) \}$ from PT-STWNC can be used to solve the marginal likelihood integral in  (\ref{eq:marglik}), by first ordering the samples with respect to $\tau$, and solving the integral numerically,

\begin{eqnarray}
\log{P(\boldsymbol{Y} \mid \text{M}_{j})}= \int\limits_{0}^{1} \E_{\boldsymbol{\theta} \mid \boldsymbol{Y}, \tau, \text{M}_{j}}[\log(P(\boldsymbol{Y} \mid \boldsymbol{\theta},\text{M}_{j} ))] d\tau  \nonumber \\ 
\approx \sum_{t=2}^{\text{T}}\Delta\tau_{t}\E_{t,{\text{M}_{j}}} \label{eq:MLPT-STWNC},
\end{eqnarray}
where $\E_{t,\text{M}_{j}}=\E_{\boldsymbol{\theta} \mid \boldsymbol{Y},\tau_{t},{\text{M}_{j}}}[\log(P( \boldsymbol{Y} \mid \boldsymbol{\theta},\text{M}_{j} ))]$ and $\Delta\tau_{t}= \tau_{t}-\tau_{t-1}$.

The variance of the marginal likelihood estimator in (\ref{eq:MLPT-STWNC}) asymptotically disappears with the number of samples, $\text{N}$. 
The variance estimator is:
\begin{equation} 
\mathrm{Var}(\log{P(\boldsymbol{Y} \mid \text{M}_{j})}) \approx \sum_{t=2}^{\text{T}} \mathrm{Var}( \Delta\tau_{t}\E_{t,{\text{M}_{j}}} ).
\end{equation}
If all the samples are unique then $\text{T}=\text{N}$ and  $\Delta\tau_{t} \approx \frac{1}{\text{N}}$ giving 
\begin{equation}
\mathrm{Var}(\log{P(\boldsymbol{Y} \mid \text{M}_{j})}) \approx  \sum_{t=2}^{\text{T}}  \frac{1}{\text{N}^{2}} \mathrm{Var}(\E_{t,{\text{M}_{j}}}). 
\end{equation}
Furthermore, $\E_{t,{\text{M}_{j}}}$ is a single sample from $P(\boldsymbol{\theta} \mid \boldsymbol{Y},\tau_{t},{\text{M}_{j}})$ giving the bounds:

\begin{eqnarray}
  \frac{1}{\text{N}} \mathrm{Var}(\boldsymbol{\theta} \mid \boldsymbol{Y},\tau=0,{\text{M}_{j}})
& > & \mathrm{Var}(\log{\text{P}(\boldsymbol{Y} \mid \text{M}_{j})}) \nonumber \\ 
& > & \frac{1}{\text{N}} \mathrm{Var}(\boldsymbol{\theta} \mid \boldsymbol{Y},\tau=1,{\text{M}_{j}}) \nonumber, \\
\end{eqnarray}
where upper and lower bounds go to zero as $\text{N}\rightarrow \infty$.

The TI via PT relies on two layers of approximation to produce marginal likelihood estimates. The first layer of approximation in TI via PT corresponds to the MCMC integration (i.e., obtaining PT samples), and the second layer occurs when  $\tau$ is  numerically integrated out from the thermodynamic integrals in (\ref{eq:discrMargPF}) and (\ref{eq:discrMarg}). Similarly, in the TI via PT-STWNC, the first approximation layer corresponds to obtaining samples from PT-STWNC, and the second layer of approximation is a result of solving the marginal likelihood integral in (\ref{eq:MLPT-STWNC}) numerically. However, since the $\tau$ is continuous, the TI via PT-STWNC removes the need for optimal temperature discretization schedule.  
Evaluations of the log likelihood can be saved within Algorithm \ref{alg:STWNC} prior to applying the temperature and consequently PT-STWNC provides marginal likelihood estimate with negligible additional computational cost.

\section{Example: Bimodal model} \label{sec:Example1}

The likelihood is bimodal with respect to $\mu$, but is unimodal with respect to  $\boldsymbol{Y}$,
\begin{equation}  P(\boldsymbol{Y} \mid \mu,\sigma^{2}) = \mbox{N}(|\mu|, \sigma^{2}), \label{eq:llik} \end{equation} and the posterior distribution of P($\mu \mid \boldsymbol{Y},\sigma^{2}$) is bimodal. The data were simulated from (\ref{eq:llik}) with $n=25$, $\mu=1.5$ and $\sigma^2=1$.

The sampling distribution of the PT-STWNC is the tempered joint posterior distribution,
\[P(\mu, \sigma^2,\tau\mid \boldsymbol{Y})\propto P(\boldsymbol{Y} \mid \mu, \sigma^2)^\tau P(\tau)P(\mu)P(\sigma^2).\]

Conjugate priors on $\mu$ and $\sigma^{2}$ were assigned,
$P(\mu) \sim \mbox{N}(0,1)$;   $ P(\sigma^2)  \sim  \mbox{InverseGamma}(1,1)$.

\subsection{Results}

\begin{figure}[!h]
	\vspace{.000003in}
    \centering\includegraphics[scale=0.16]{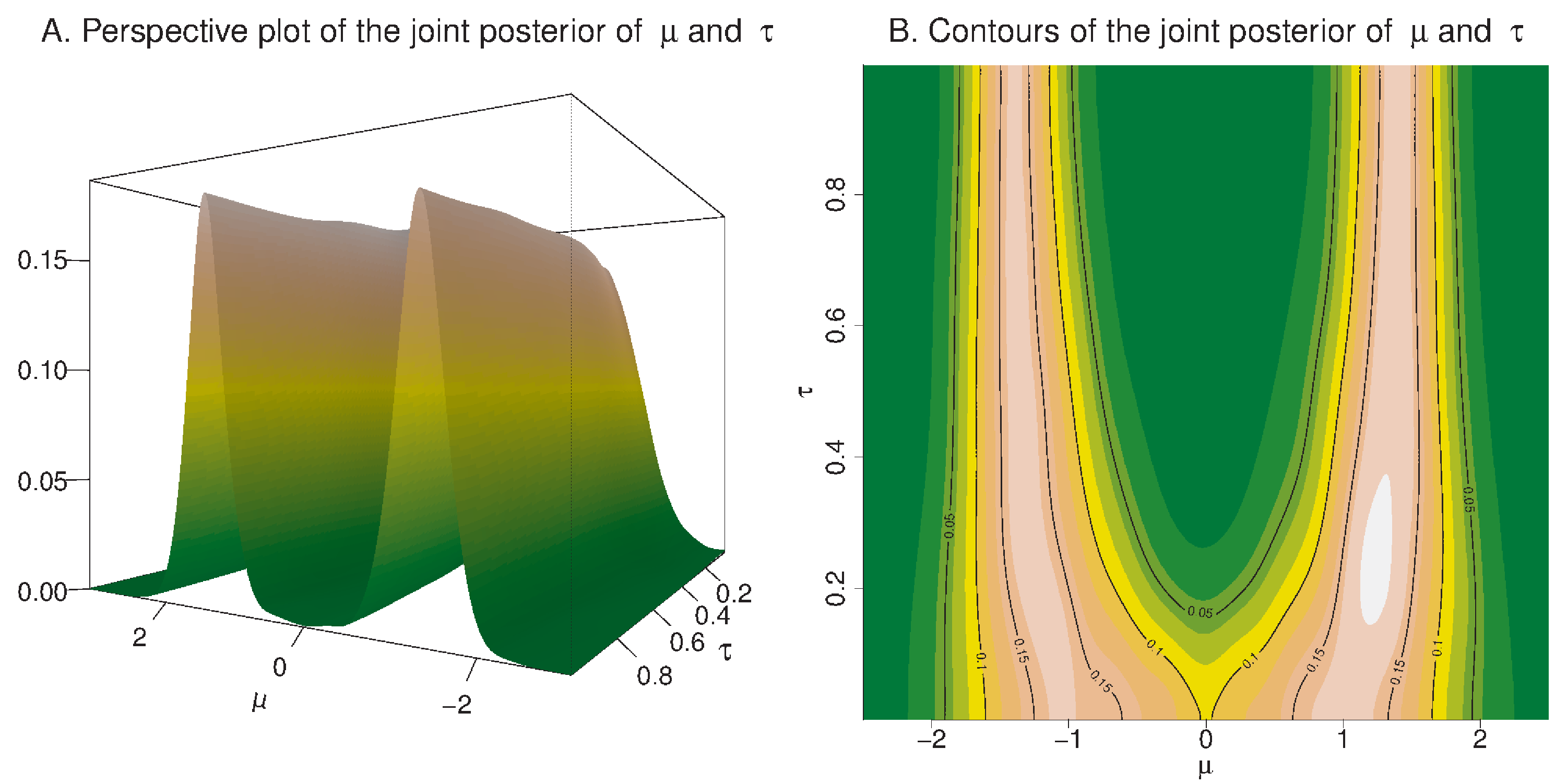}  
	\vspace{.000003in}
	\caption{The bimodal model --  A). perspective plot of the joint  posterior distribution of $\mu \mbox{ and } \tau$, and B). the corresponding contour plot. For illustration purpose, the posterior distribution in this plot was obtained from the one parameter model, where the parameter $\mu$ was sampled, while $\sigma^{2}$ was treated as known and fixed to the true value of 1 }
	\label{fig:Trg}
\end{figure}

The PT-STWNC algorithm  was run for 50,000 iterations  with the first 15,000 samples discarded as burn-in. 
Figure~\ref{fig:Trg} shows the sampled joint posterior distribution of $\mu$ and $\tau$ obtained from the PT-STWNC `tempered' chain.
The perspective and contour plots in Figure~\ref{fig:Trg} illustrate uniform profile as seen in the  two ridges of the posterior surface having approximately constant maximum height along the $\tau$ axis. The last observation is a direct consequence of the constraint that the profile posterior distribution of $\tau$ while profiling over $\mu$ is uniform on the interval [0,1]. 

\begin{figure}[!h]
	\centerline{\includegraphics[scale=0.30]{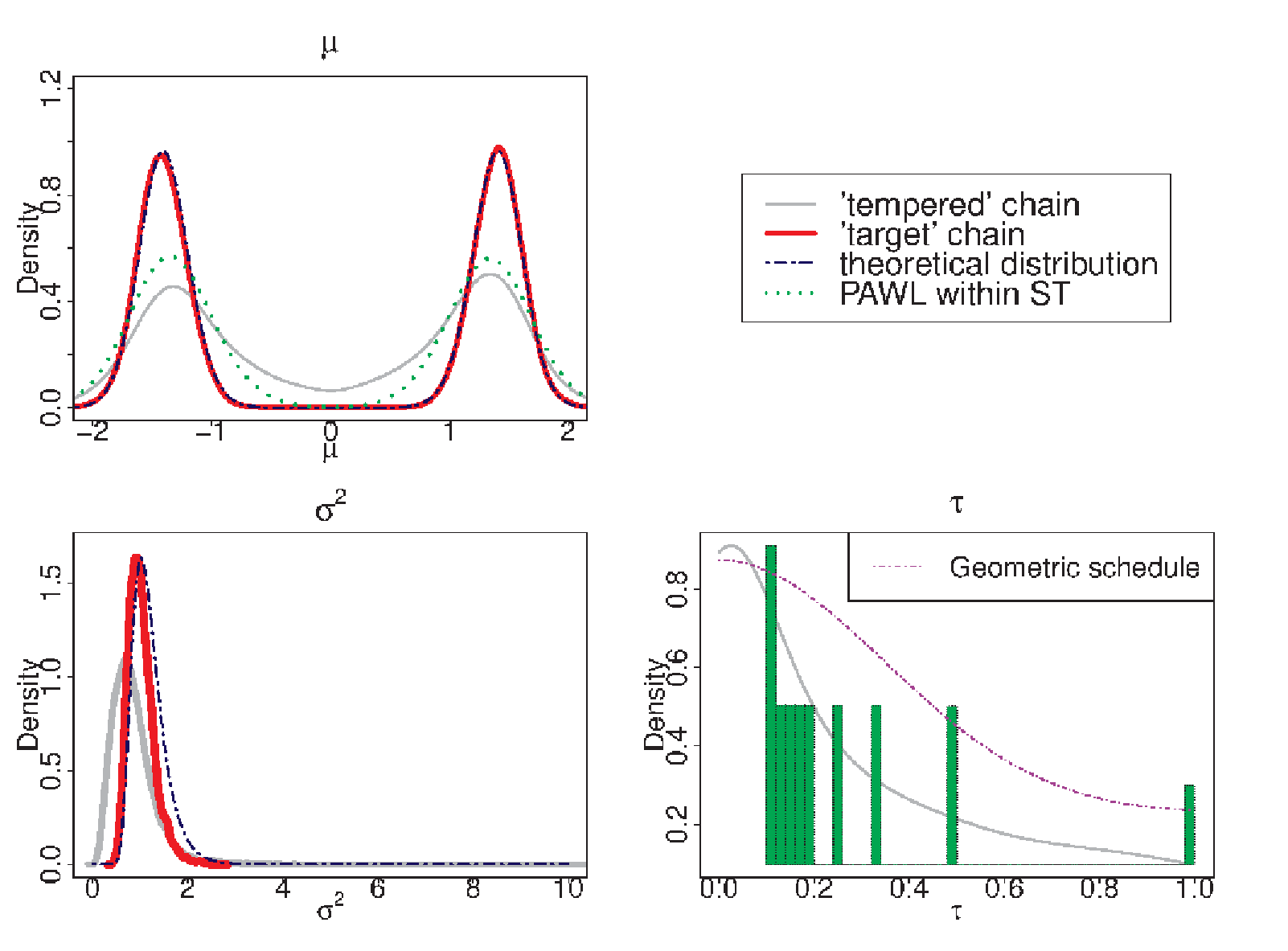}}
	\vspace{.00003in}
	\caption{The bimodal model -- marginal posterior distributions of $\mu,\tau \mbox{ and } \sigma^{2}$ from the two parameter model. Distributions in gray, red and blue color correspond to samples obtained from the PT-STWNC `tempered' chain, the PT-STWNC `target' chain and the theoretical distribution, respectively. The green color corresponds to samples from the PAWL within ST algorithm. Bars in the histogram of $\tau$ correspond to the marginal distribution of the discrete temperature obtained from the PAWL within ST}
	\label{fig:pairwisePlots}
\end{figure}

The wide contours of Figure~\ref{fig:Trg}B at low values of $\tau$ demonstrate that the 'tempered' chain spends a lot of time sampling at low values of the inverse temperature, thus taking large steps to move between the two modes. Similarly, the marginal distribution of sampled $\tau$  demonstrates that low values of $\tau$ are sampled  with higher frequencies (see Figure~\ref{fig:pairwisePlots}, gray color).
The PT-STWNC `target' chain  updates the parameters of interest at $\tau=1$, which results in drawing samples from the target posterior distribution (Figure~\ref{fig:pairwisePlots}, diagonal, blue color). 

PAWL within ST was also run on this toy example and the marginal distributions of $\mu$ and $\tau$ were compared to those from the PT-STWNC. The plot of the marginal distribution of $\mu$  is similar to the marginal distribution of $\mu$ obtained from the 'tempered' PT-STWNC chain (Figure~\ref{fig:pairwisePlots}, green color). The marginal distribution of the discrete $\tau$ from the PAWL within ST has similar shape as that of the continuous $\tau$ from the PT-STWNC. 
Table~\ref{tbl:estim} shows the posterior means of  $\mu$ and $\sigma^{2}$ compared to their theoretical values. Full implementation details are given in the Appendix \ref{ImplementDetails}.

\begin{table}[!h]
	\caption{Parameter estimates v.s. theoretical values} \label{tbl:estim}
	\small
	\centering
	\begin{tabular}{rlllll}
		\hline
		&   Parameter estimates & Theoretical results  \\ 
		\hline
				$\mu$  & 1.405(0.0019); -1.413(0.0019) & 1.409 ; -1.409 \\ 
		$\sigma^{2}$  &  1.026(0.015)     &  1.148   \\
		\hline
	\end{tabular}
	\caption*{The bimodal model -- estimated posterior means (from the 'target' chain) and theoretical posterior means (from the target distribution) for each of the two modes of $\mu$ and for $\sigma^2$ from the two parameter model. Monte Carlo errors of the point estimates, obtained as per \cite{doi:10.1146/annurev-statistics-022513-115540}, are given in brackets}
\end{table}

\subsection{Marginal likelihood estimation}  \label{sec:mllik_exmpl1}

The PT-STWNC  was used to estimate the marginal likelihood of the bimodal model introduced in the Section \ref{sec:Example1}. 
The PT-STWNC marginal likelihood estimates were compared to the following three approaches: $i).$ analytical solution to the marginal likelihood integral in (\ref{eq:ordmllik}), $ii).$ the thermodynamic integration via PT with bias correction by \cite{calderhead2009estimating} (TI-PT-B) and $iii).$ the thermodynamic integration via PT without bias correction by \cite{friel2008marginal} (TI-PT-NB).

This is a toy example and one of the very few cases where a closed form of the marginal likelihood integral exists (see Appendix \ref{sec:AppendixA}). The PT-STWNC marginal likelihood estimates were obtained directly from the 'tempered' and 'target' chain using the equation (\ref{eq:MLPT-STWNC}).
TI via PT estimates were obtained by running PT with T=30 chains. The inverse temperature schedule was chosen as the geometric schedule, $t_{i}=(\frac{i}{\text{T}})^5$.

The  TI-PT-NB and the TI-PT-B estimates in  Table~\ref{tbl:pants}, were obtained using samples from the same run of the PT. The only difference between the TI-PT-NB and the TI-PT-B is the bias term as per equation  (\ref{eq:discrMarg}). The results  in Table~\ref{tbl:pants} show that the thermodynamic integral bias term  has a small effect on the marginal likelihood estimate.   Table~\ref{tbl:pants}  also found that increased number of PT chains (T=60 and T=100) did not yield substantially better TI-PT-NB and TI-PT-B estimates. This result complies with the finding by \cite{Ahlers2008}, who demonstrated that while thermodynamic integration via PT performs well in unimodal case, the method exhibits substantiative bias in a bimodal case with a tractable marginal likelihood. The results from PT and our PT-STWNC both exhibit bias, this is consistent with \cite{Ahlers2008}, who also found that the estimates do not improve with increased number of parallel chains. Those authors trace back this problem to the incomplete equilibrium between the two modes, which leads to failure to reproduce the exact mixture probabilities. In addition, the bias in PT-STWNC, TI-PT-NB, and TI-PT-B did not reduce when the integral in (\ref{eq:MLPT-STWNC}) was obtained by combining all the samples from the 20 replicate runs.

\begin{table*}[!h]
	\caption{Marginal log-likelihood of the bimodal model} \label{tbl:pants}
	\small
	\centering
	\begin{tabular}{rrlllll}
	\specialrule{.05em}{1em}{0em} 
		& & Analytic solution &  PT-STWNC  & TI-PT-B  & TI-PT-NB \\ 
		\specialrule{.05em}{.05em}{.05em} 
		& & -38.946 & -37.767(0.019) &   &  \\
		\specialrule{.0002em}{.1em}{.1em} 
		&\text{T} =30 &   &   & -38.254(0.006) & -38.261(0.006)\\ 
		&\text{T} =60 &- & - & -38.247(0.004) &  -38.254(0.005)\\ 
		&\text{T} =100 &- & -   &  -38.247(0.004) & -38.254(0.003)\\
		\specialrule{.05em}{0em}{0em}
	\end{tabular}
	\vspace{.2in}
	\caption*{The bimodal model -- marginal log-likelihood estimates from: analytic solution, PT-STWNC, thermodynamic integration via PT with bias correction \citep{calderhead2009estimating} (TI-PT-B) and thermodynamic integration via PT without bias correction \citep{friel2008marginal} (TI-PT-NB). The PT-STWNC, the TI-PT-B and the TI-PT-NB estimates are based on 20 independent runs. Standard deviations of the marginal likelihood estimates obtained from the $20$ runs are given in brackets. The TI-PT-B and the TI-PT-NB estimates with $\text{T}=60$ and $\text{T}=100$ chains are also obtained from $20$ independent runs. All the marginal log-likelihood estimates were obtained using the one parameter model, where the parameter $\mu$ was sampled, while $\sigma^{2}$ was fixed to the true value of 1. Details on the convergence of the PT chains are given in the Appendix~ \ref{ImplementDetails}}
\end{table*}

\section{Example: Galaxy data} \label{sec:Galaxy}

The Galaxy data comprises velocities of $82$ galaxies that diverge from our galaxy studied by \cite{postman1986probes, 1995, neal1999erroneous}. The data are univariate identically and independently distributed samples from mixture of $K$ Gaussian components and denoted as $\boldsymbol{Y}=(y_{1},y_{2},..,y_{n})^{'}$, with $n=82$. Parameters of the model are given as $\boldsymbol{\theta}=\left(\boldsymbol{\mu}, \boldsymbol{\sigma^{2}} ,\boldsymbol{p} \right)^{'}$ where  $\boldsymbol{\mu}=\left(\mu_{1},..,\mu_{K}\right)^{'}$ is a vector of mixture component means, $\boldsymbol{\sigma^{2}}=\left(\sigma^{2}_{1},..,\sigma^{2}_{\text{K}}\right)^{'}$ is  a vector of mixture component variances and $\boldsymbol{p}=\left(p_{1},..,p_{K}\right)^{'}$ is a vector of mixture probabilities. The $k$-th mixture component has distribution $\mbox{N}( \textbf{Y} \mid \mu_{k},\sigma_{k}^2 )$.
Then likelihood function is

\begin{equation}
P(  \textbf{Y} \mid \boldsymbol{\mu},\boldsymbol{\sigma^{2}}, \boldsymbol{p} )= \prod_{i=1}^{n}\sum_{k=1}^{K} p_{k} \mbox{N}(y_{i} \mid \mu_{k},\sigma_{k}^2).  
\end{equation}
Conjugate priors were assigned,
$P(\mu_{k})  \sim  \mbox{N}(20, 100), $
$P(\sigma_{k}^2)  \sim  \mbox{IGamma}(shape=3, scale=20) $ and
$P(p_{1},..,p_{K})  \sim \mbox{Dirichlet}(\alpha_{1}=1,..,\alpha_{K}=1)$.

A latent variable \textbf{Z} such that $P(Z_{ik}=1\mid p_{k})=p_{k}$  was introduced to help derive the necessary sampling  distributions. \textbf{Z} is a $n \times K$ matrix of indicator variables in which $\text{Z}_{i,k}=1$ indicates the data point $i$ belongs to the mixture component $k$. Then likelihood of data point $\{y_{i}\}$ conditional on $\text{Z}_{ik}$ is,
\begin{equation*}
P(y_{i} \mid   \text{Z}_{ik}=1,\boldsymbol{\mu},\boldsymbol{\sigma^{2}},\boldsymbol{p}  )=\text{N}(y_{i} \mid    \mu_{k},\sigma_{k}^{2} ), \nonumber \\
\end{equation*}
and the joint distribution  of $\{ y_{i}\}$ and  $Z_{ik}$ is,
\begin{equation*}
P(y_{i},Z_{ik}=1 \mid   \boldsymbol{\mu},\boldsymbol{\sigma^{2}},\boldsymbol{p}  )=p_{k}N(y_{i} \mid    \mu_{k},\sigma_{k}^{2} ). \nonumber \\
\end{equation*}
Each row of \textbf{Z}  is a multinomial variable with probabilities,
\begin{eqnarray}
P(\text{Z}_{ik}=1 \mid y_{i},\boldsymbol{\mu},\boldsymbol{\sigma^{2}},\boldsymbol{p} )=\frac{p_{k}\mbox{N}(y_{i} \mid \mu_{k},\sigma_{k}^{2}   )}{\sum_{j=1}^{K} p_{j}\mbox{N}(y_{i} \mid \mu_{j},\sigma_{j}^{2}   ).} \label{eq:Z}
\end{eqnarray}

The joint posterior distribution of the parameters of interest $\boldsymbol{\theta}=\left(\boldsymbol{\mu},\boldsymbol{\sigma^2},\boldsymbol{p}\right)^{'}$, latent variable $\textbf{Z}$ and the inverse temperature parameter $\tau$ is,

\small
\begin{eqnarray*}
	P(\boldsymbol{\mu},\boldsymbol{\sigma^2},\boldsymbol{p},\boldsymbol{Z},\tau \mid \textbf{Y}) & \propto & 
	\left(\prod_{i=1}^{n}\sum_{k=1}^{K}  P(y_{i} \mid   \text{Z}_{ik}=1,\boldsymbol{\mu},\boldsymbol{\sigma^2},\boldsymbol{p}) \right) ^{\tau}   \times   \\
	& &  P_{\boldsymbol{p}}(\boldsymbol{p}) \left(  \prod_{k=1}^{K} P_{\mu_{k}}(\mu_{k}) \right)\left(  \prod_{k=1}^{K}P_{\sigma_{k}^2}(\sigma_{k}^2) \right) \times   \\
	& & \left(  \prod_{i=1}^{n}\sum_{k=1}^{K} P(\mbox{Z}_{ik} \mid p_{k})  \right) P(\tau).  \label{jointpost}
\end{eqnarray*}
\normalsize

\subsection{Results}

\begin{figure}[!h]
	\centerline{\includegraphics[scale=0.32]{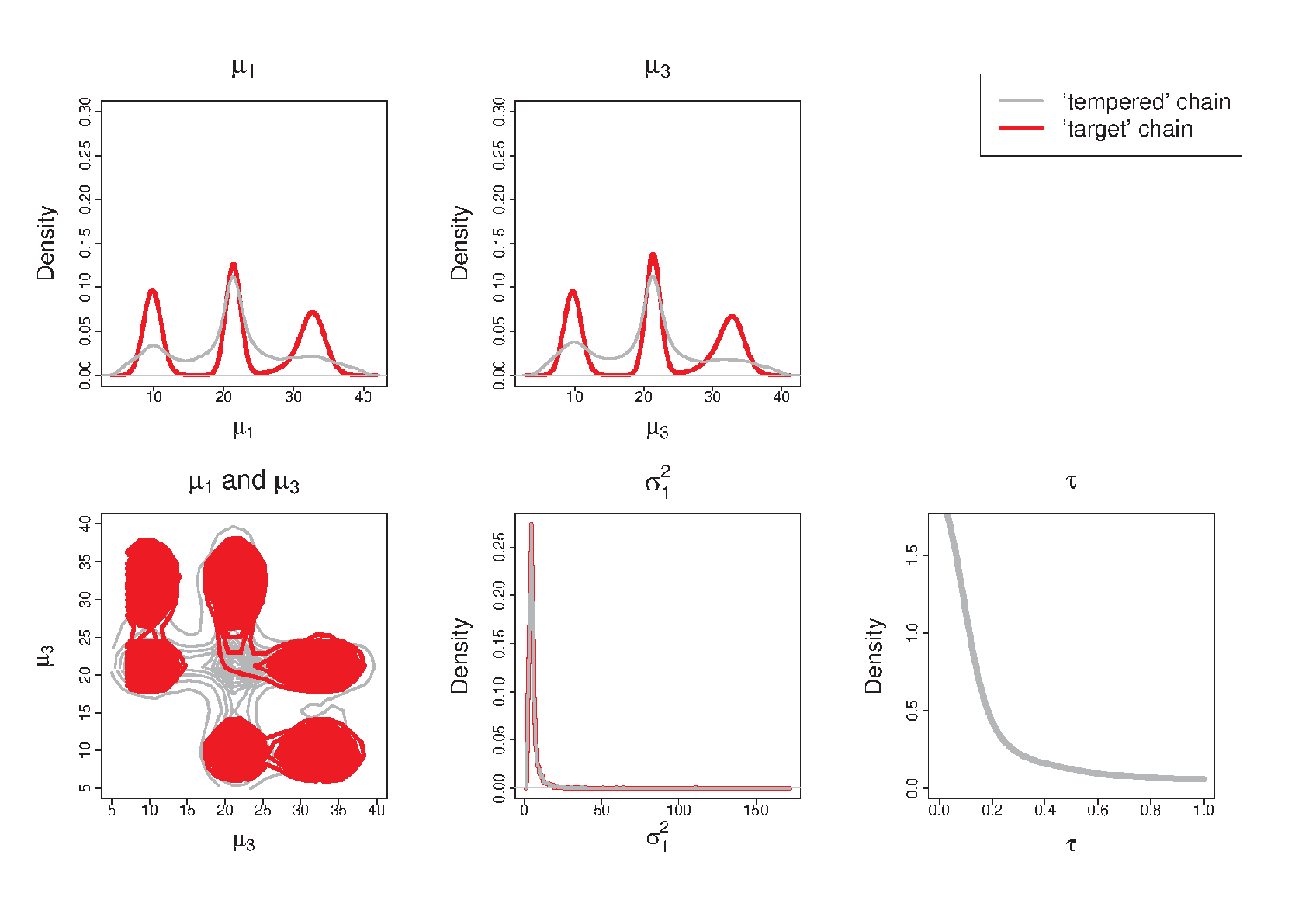}}
	\vspace{.003in}
	\caption{Galaxy data, model with three components unequal variances -- marginal  posterior distributions of the sampled parameters $\mu_{1},\mu_{3},\sigma^2_{1},\tau$ and bivariate plot of $(\mu_{1},\mu_{3})$. Gray and red color correspond to samples obtained from the `tempered' and `target' chain, respectively. 
	}
	\label{fig:GalaxyPlot}
\end{figure}

The PT-STWNC algorithm  was run  for 35,000 iterations  with the first 1000 generated samples being removed as burn-in. Multi-modality in the Galaxy data with three components is illustrated by marginal distributions of $\mu_{1},\mu_{3},\sigma^2_{1} \mbox{ and } \tau$ and bivariate joint posterior distribution of $(\mu_{1},\mu_{3})$  (Figure~\ref{fig:GalaxyPlot}). Plot of the marginal distribution of $\tau$  shows that $\tau$ spends a lot of time at values close to zero, which is crucial for the sampler to move easily between the isolated modes and produce accurate estimates (Figure ~\ref{fig:GalaxyPlot}).

\subsection{Marginal likelihood estimation for the Galaxy data}

The PT-STWNC  was used to perform model selection on the Galaxy data set via the choice of $K$. The marginal likelihood  was calculated for the Galaxy data to compare the following models: two components with equal variances, three components with unequal variances, three components with equal variances, four components with unequal variances and five components with unequal variances. A closed form expression of the marginal likelihood integral is not available.

The two thermodynamic integration approaches from the Section \ref{sec:TI}, were used to compare model selection abilities with PT-STWNC. Log marginal likelihood estimates and Bayes Factors demonstrate that all three estimation techniques agree that the model with five components is the best model (Tables~\ref{tbl:Galaxymllik} and \ref{tbl:BayesFactors}). In addition, all the three estimation methods find support for the models with $3-5$ components, while the worst model is the model with $2$ components and equal variances. Findings from our model selection study comply with the results from previous studies. For instance, \cite{steele2010performance} argued that the number of components in the Galaxy data is not known, and concluded that the models with $3$ and $6$ components are reasonable fit to the data.  Gibbs sampling was used to show that the model with $3$ components is the best model \cite{chib1995marginal}. Evolutionary Monte Carlo was used in combination with bridge sampling 
to demonstrate that the model with 5 components is the best model and $2$ components with equal variances is the worst model  \cite{liang2001real}. In addition, \cite{liang2001real} found support for the models with $3-5$ components. Reversible Jump MCMC (RJMCMC) was used to conclude that the models with $5-7$ components are good fit to the data  \cite{richardson1997bayesian}.

The  TI-PT-NB and the TI-PT-B estimates in  Table~\ref{tbl:Galaxymllik} were obtained using samples from the same run of the PT. Hence, the only difference between the TI-PT-NB and the TI-PT-B is in the bias term as per equation  (\ref{eq:discrMarg}). The results  in the Table~\ref{tbl:Galaxymllik} show that the  estimates from the TI-PT-NB and the TI-PT-B are nearly the same, which suggests that the thermodynamic integral bias term has near zero effect in this example.

\begin{table*}[!h]
 
	\caption{Log marginal likelihood estimates of Galaxy data} \label{tbl:Galaxymllik}
	\centering
    {\begin{tabular}{llllllll}
   			\hline
			Model fitted   &   PT-STWNC     &  TI-PT-B  &      TI-PT-NB    \\
	\specialrule{.05em}{.05em}{.05em}
			1. 2 components equal variances & -241.99(0.25) & -238.02(0.03) &  -238.03(0.02)  \\ 
			2. 3 components unequal variances  &  -228.67(0.39) & -224.26(0.03) &  -224.28(0.03)    \\ 
			3. 3 components equal variances    & -236.10(0.47) & -224.20(0.06)  & -224.23(0.05)      \\ 
			4. 4 components unequal variances   & -222.72(0.28) &  -223.88(0.02)  &  -223.89(0.02)        \\ 
			5. 5 components unequal variances   &  -221.62(0.50)  &  -223.85(0.02)  &  -223.85(0.02)  \\  
			\hline
	\end{tabular}}
	\caption*{Galaxy data -- log marginal likelihood estimates obtained from the PT-STWNC, the TI-PT-B and the TI-PT-NB. Equations (\ref{eq:MLPT-STWNC}), (\ref{eq:discrMarg}) and (\ref{eq:discrMargPF})  were used to obtain marginal likelihood estimates from the PT-STWNC, the TI-PT-B and the TI-PT-NB, respectively,  for each of the five different models. The TI-PT-B and the TI-PT-NB estimates were obtained from $\text{T}=30$ PT chains  using a geometric temperature schedule that tempers towards the prior $t_{i}=(\frac{i}{\text{T}})^5$. All the marginal likelihood  estimates were obtained from $10$ independent runs of each of the estimation techniques for each of the five models. Standard deviations of the marginal likelihood estimates from 10 runs are given in brackets}
\end{table*}

\begin{table}[!h]
	\caption{ log Bayes factors}
	\label{tbl:BayesFactors}
	\centering
	\begin{tabular}{rrrr}
		\hline
		& PT-STWNC & TI-PT-NB & TI-PT-B \\ 
		\hline
		$\log{BF21}$ & 13.32 & 13.74 & 13.75 \\ 
		$\log{BF31}$ & 5.89 & 13.79 & 13.82 \\ 
		$\log{BF41}$ & 19.27 & 14.14 & 14.13 \\ 
		$\log{BF51}$ & 20.37 & 14.17 & 14.16 \\ 
		$\log{BF32}$ & -7.44 & 0.05 & 0.06 \\ 
		$\log{BF42}$ & 5.85 & 0.39 & 0.38 \\ 
		$\log{BF52}$ & 7.05 & 0.43 & 0.41 \\ 
		$\log{BF43}$ & 13.381 & 0.34 & 0.31 \\ 
		$\log{BF53}$ & 14.49 & 0.37 & 0.34 \\ 
		$\log{BF54}$ & 1.1 & 0.03 & 0.02 \\ 
		\hline
	\end{tabular}
	\caption*{Bayes Factors obtained by applying the equation (\ref{eq:BF_1}) to the log marginal likelihood estimates in the Table~\ref{tbl:Galaxymllik}}   
\end{table}

\section{Example: Susceptible-Infected-Recovered (SIR) epidemiological model with real data } \label{sec:SIR}

We illustrate the PT-STWNC on a Susceptible-Infected-Recovered (SIR) epidemiological model for number of daily deaths due to the black plague. The data were collected by the grave digger during the second  black plague outbreak in the village of Eyam, UK, from June 19, 1666 to November 1, 1666 \citep{massad2004eyam}. The village had quarantined itself to avoid spreading the disease to the neighboring villages. Therefore,  the population size is fixed to $\text{N}=261$, and the population  is stratified into groups of susceptible $\text{S}(t)$, infected  $\text{I}(t)$ and removed $\text{R}(t)$ individuals, $\text{N}=\text{S}(t)+\text{I}(t)+\text{R}(t)$. Since there is no recovery from the plague, the number of deaths correspond to the number of removed individuals up to time $t$, $\text{R}(t)$       \citep{campbell2014anova,golchi2016sequentially}

The disease spread dynamics can be described by the system of ordinary differential equations (ODE),
\begin{align}
\label{eq:SIR}
\frac{d\text{S}}{dt}   =-\beta \text{S}(t)\text{I}(t), \mbox{ }  \frac{d\text{I}}{dt}   = \beta \text{S}(t)\text{I}(t)-\alpha I(t), \mbox{ } \frac{d\text{R}}{dt}   & = \alpha \text{I}(t)
\end{align}
where $\alpha$  describes the rate of death once the individual is infected and $\beta$ describes the plague transmission. Additional to the model parameters $\boldsymbol{\theta} =\left(\alpha, \beta\right)^{'}$, the ordinary differential equations model requires estimates of the initial states $(\text{S}(0),\text{I}(0),\text{R}(0))^{'}$. At the initial time the population consists of susceptible and infected individuals, and therefore  $\text{R}(0)=0$ and $\text{S}(0)=\text{N}-\text{I}(0)$. Consequently, the only initial state parameter is $\text{I}(0)$ so that the unknown parameters of the model are $\boldsymbol{\theta}=\left(\alpha,\beta,\text{I}(0)\right)^{'}$. The data denoted as $\boldsymbol{Y}= (y_{1},.., y_{n})^{'}$ with $n=136$, represent cumulative number of deaths up to times $t_{1},..,t_{n}$. The data points $\boldsymbol{Y}$ were modeled by a Binomial distribution with expected value equal to the solution to the system (\ref{eq:SIR}), $R_{(\alpha,\beta,\text{I}(0))}(t)$, where $t \in \{t_{1},..,t_{n}\}$ .

The number of susceptible $\text{S}(t)$ and infected $\text{I}(t)$ are not observed. However, the number of infected at the end of the plague is 0,  and the number of infected at time one before the end of the plague must therefore equal 1 \citep{campbell2014anova}. Two additional data points on number of infected individuals $\boldsymbol{X}=(x_{n-1}=1,x_{n}=0)^{'}$ at times $( t_{n-1},t_{n})^{'}$ were modeled using Binomial distribution with expected value equal to $\text{I}_{(\alpha,\beta,\text{I}(0))}(t)$ for $t \in (t_{n-1},t_{n})^{'}$,
\begin{eqnarray}
P( \boldsymbol{Y} \mid \alpha,\beta,\text{I}(0) ) & = & \prod_{i=1}^{n}    \mbox{Binomial} \bigg(y_{i} \mid N, \frac{\text{R}_{(\alpha,\beta,\text{I}(0))}(t_{i})}{\text{N}} \bigg) \times \nonumber \\  
& &\prod_{i=n-1}^{n} \mbox{Binomial}  \bigg(x_{i} \mid  \text{N}, \frac{\text{I}_{(\alpha,\beta,I(0))}(t_{i})}{\text{N}} \bigg) \nonumber \\
\end{eqnarray}

Prior distributions for 
$\boldsymbol{\theta}=\left(\alpha,\beta,\text{I}(0) \right)^{'}$ 
were chosen to be:
$\alpha,\beta \sim  \mbox{Gamma}(1,1), $
$\text{I}(0) \sim   \mbox{Binomial}(\text{N},\frac{5}{\text{N}}). $

Parameters $\alpha$ and $\beta$ are continuous and $\text{I}(0)$ is discrete. The discrete nature of the $\text{I}(0)$ induces multi-modality in the likelihood surface. This mixture of discrete and continuous parameters in the model imposes difficulties in sampling from the posterior distribution. Standard MCMC could get easily trapped in local modes of the posterior of the parameters of interest.

\subsection{Results}

\begin{figure}[!h]
	\centerline{\includegraphics[scale=0.18]{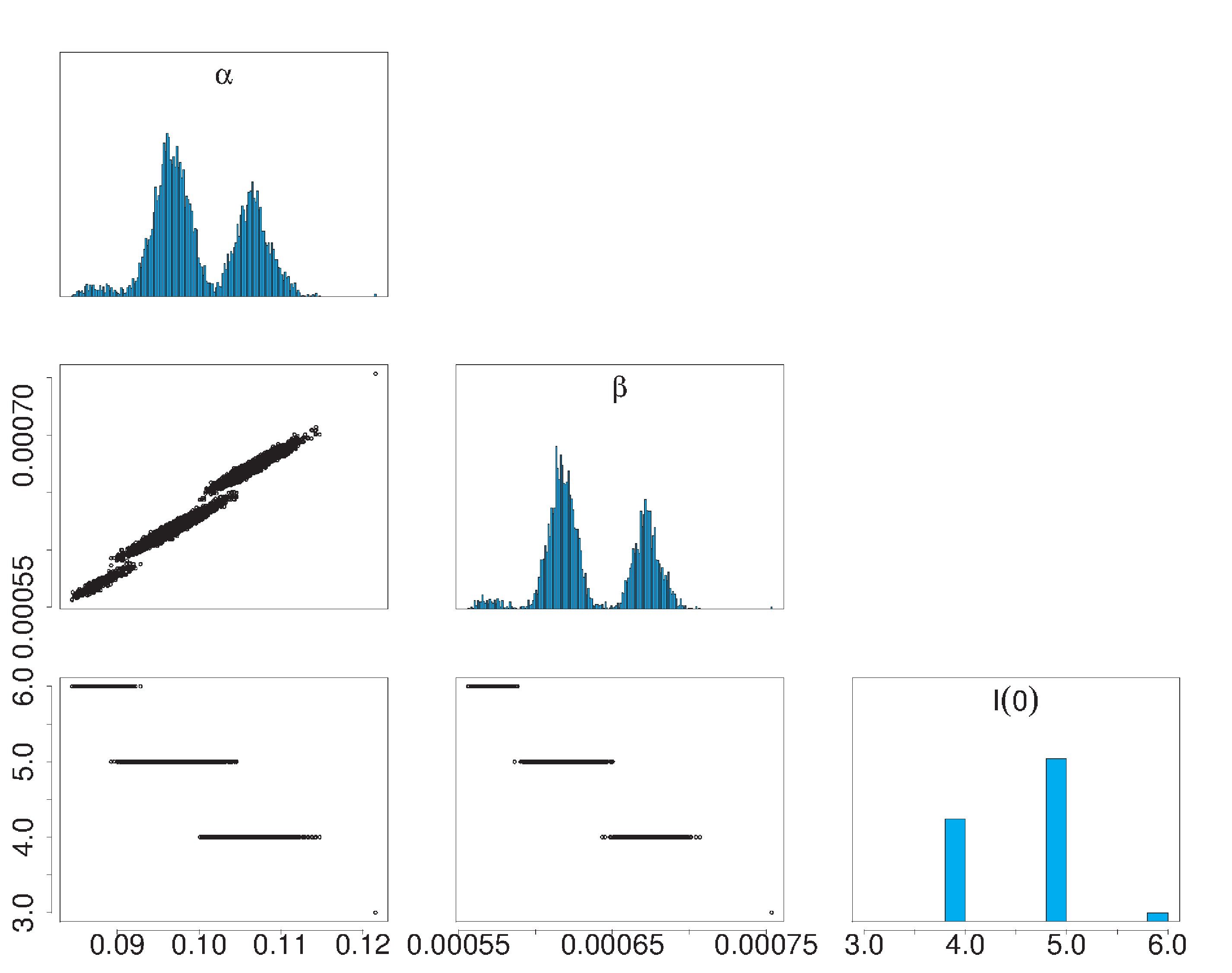}}
	\caption{SIR model -- marginal (diagonal) and bivariate joint (off-diagonal) posterior distributions of sampled parameters  $\alpha,\beta \mbox{ and } \text{I}(0)$ obtained from the `target' chain}
	\label{fig:pairsPlot_SIR}
\end{figure}

The PT-STWNC  was run on the SIR model for  35,000 iterations   with 1000 burn-in samples. We applied a conjugate-gradient optimization routine for continuous parameters in parallel conditional on discrete parameters allowing a follow-up line search over discrete parameters.   Multi-modality and topological challenges of the model are illustrated by the marginal distribution plots  (Figure~\ref{fig:pairsPlot_SIR}, diagonal) and by the bivariate joint posterior distributions plots (Figure~\ref{fig:pairsPlot_SIR}, off-diagonal) of $\boldsymbol{\theta}=\left(\alpha,\beta,\text{I}(0)\right)^{'}$.  The marginal distributions of $\alpha$ and $\beta$ exhibit structures with three isolated modes. In Figure~\ref{fig:pairsPlot_SIR} (off-diagonal), clouds in the joint posterior distribution of $\alpha$ and $\beta$ represent the modes which correspond to the discrete samples of $\text{I}(0)=\{6,5,4,3\}$ from left to right.

\begin{figure}[!h]
	\centerline{\includegraphics[scale=0.18]{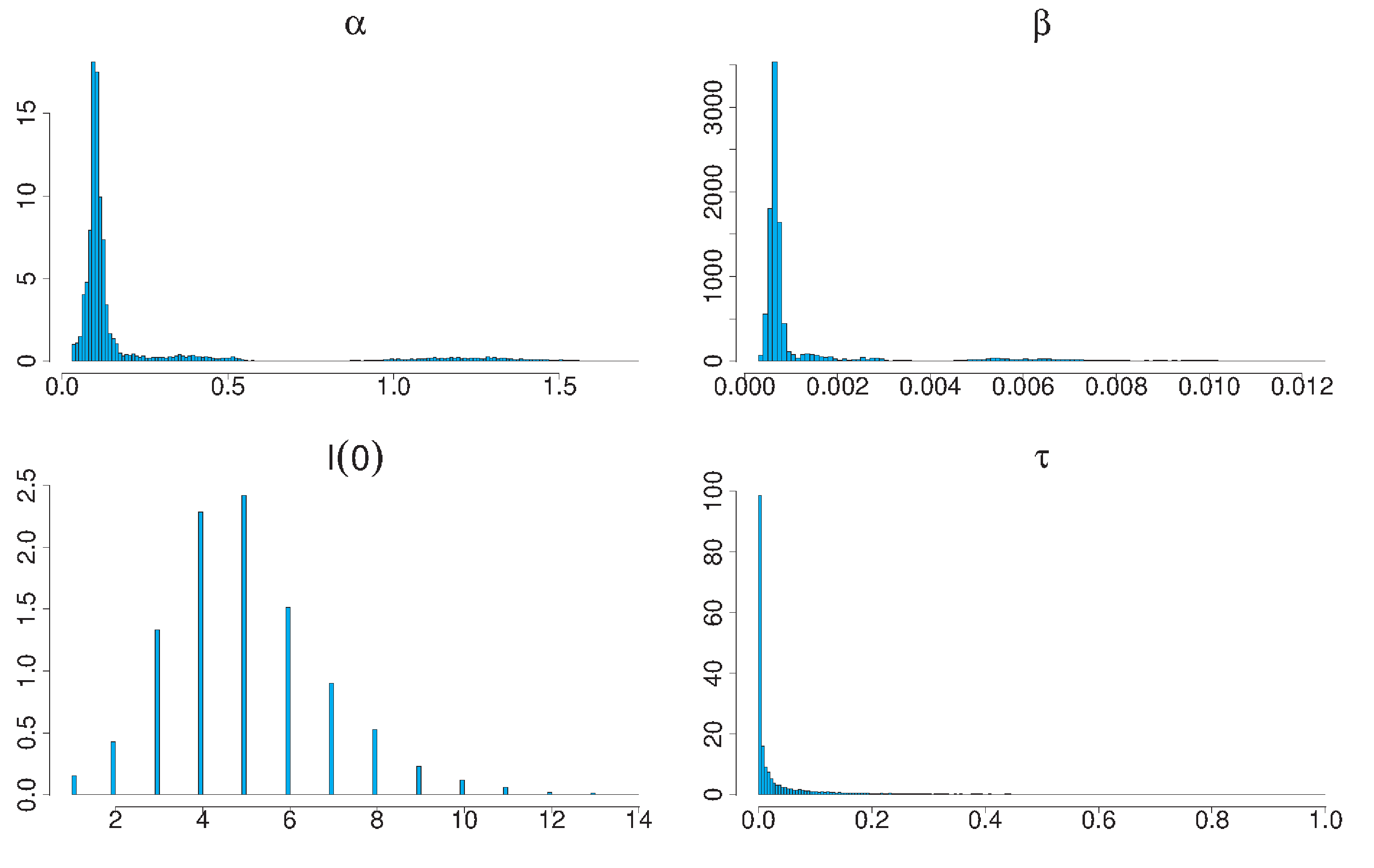}}
	\caption{The SIR model -- marginal posterior distributions of sampled parameters  $\alpha,\beta, \text{I}(0) \mbox{ and } \tau $ obtained from the `tempered' chain}
	\label{fig:pairsPlot_SIR1}
\end{figure}

\begin{figure}[!h]
 
    \centering{\begin{tabular}{@{}l@{}}
     \includegraphics[scale=0.16]{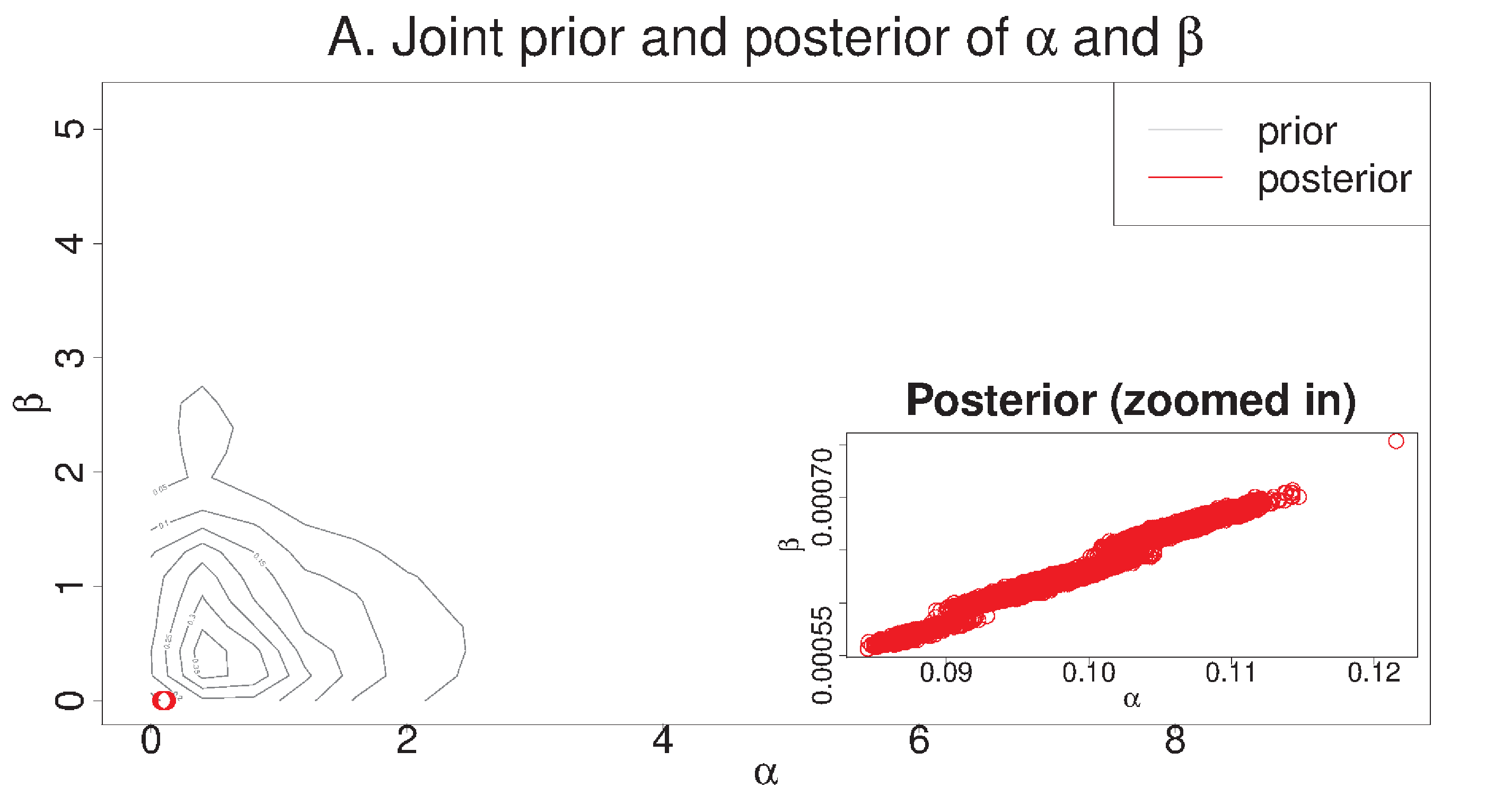}
     \end{tabular}\par}
     \vspace{0.01in}
    {\begin{tabular}{@{}r@{}}
     \includegraphics[scale=0.16]{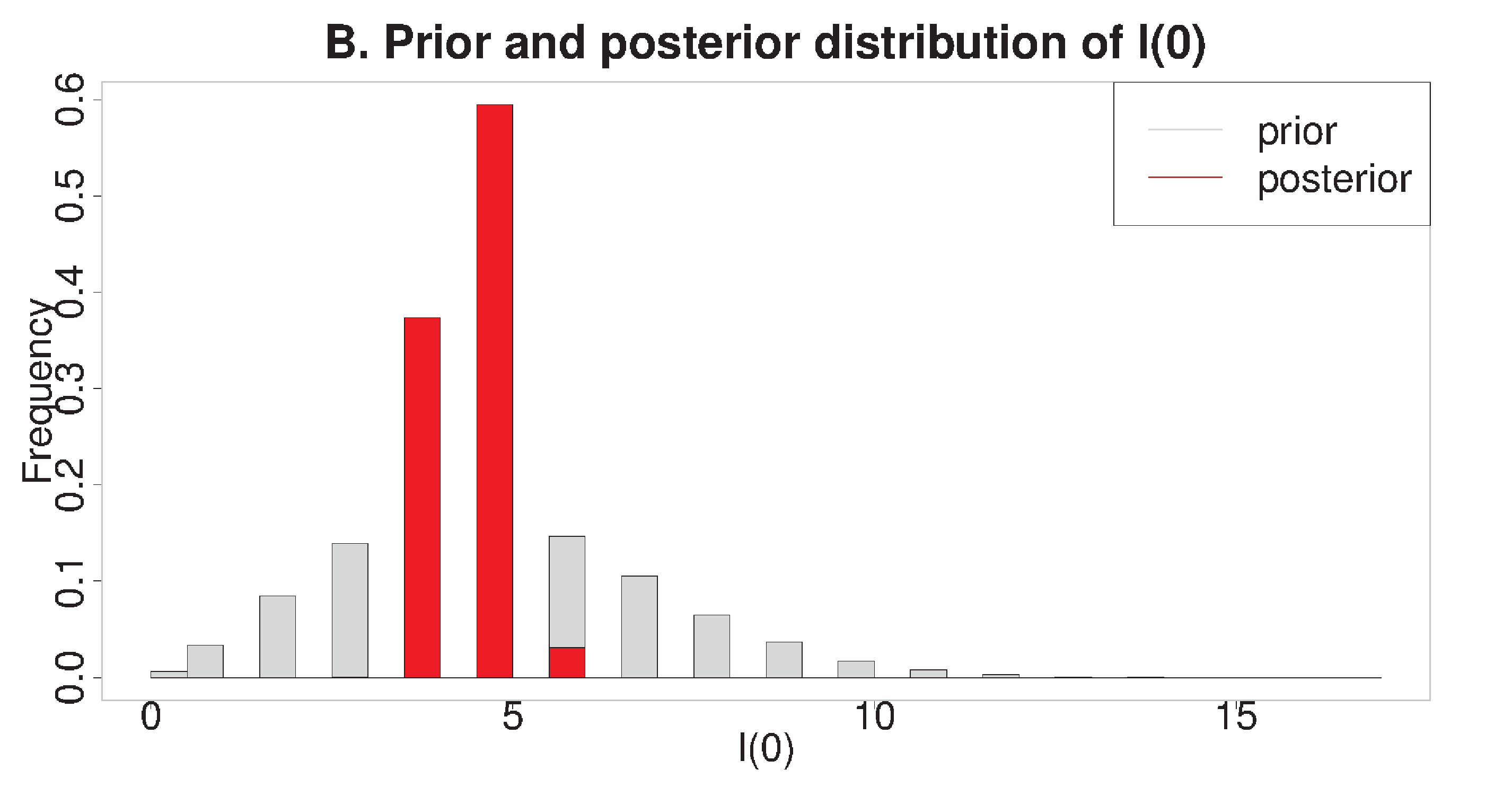} 
     \end{tabular} \par}
 	\vspace{.000003in}
	\caption{The SIR model -- parameter space of the prior (gray color) versus the posterior space (red color) of the parameters in SIR model. The plot A shows contours of the joint prior distribution of the parameters $\alpha$ and $\beta$. The small red dots close to the origin correspond to the joint posterior distribution. Parameter space of the prior and posterior distribution of the parameter $\text{I}(0)$ are shown in the plot B }
\label{fig:PriorPostAB}
\end{figure}

Histograms of the marginal distributions of $\alpha,\beta,\text{I}(0)$ and $\tau$  obtained from the `tempered' chain (Figure~\ref{fig:pairsPlot_SIR1}), illustrate the complexity and topological challenges of the model as well as the need for exploring the diffuse prior parameter space in order for PT-STWNC to draw samples from the target distribution. Figure~\ref{fig:PriorPostAB} demonstrates that the prior parameter space (the grey contour lines) is much more diffuse than that of the joint posterior distribution of $\alpha$ and $\beta$ (the red dots, which when zoomed-in assumes the shape of the target posterior distribution). 
Consequently, the algorithm spends much time sampling at near-zero values of $\tau$  thus exploring the prior parameter space. 

For implementation details we refer the reader to Appendix \ref{ImplementSIR}.

\subsection{Computational Acceleration by Approximating $P(\tau)$}\label{sec:accel}


A promising direction for accelerating the implementation is through optimal manifold approximation.  Prior to beginning MCMC iterations, optimization was performed over a grid of 301 values of $\tau$ and an interpolator is defined so as to replace the optimization within MCMC iterations with evaluation of the interpolator instead.  In this example, $\theta_{\max}(\tau)$ is optimized over a $\log_{10}$-uniform grid of 301 values of $\tau\in[10^{-15},1]$. The resulting values are then interpolated using a $4^{th}$ order b-spline basis with 60 $\log_{10}$-uniformly spaced knots across $[10^{-15},1]$.   The computational time for the initial 301 optimizations and spline interpolator set up was under one minute.  After that initial computational cost, the time per iteration of this approximation to PT-STWNC is equivalent to that of parallel tempering also operating with the same number of chains.  This is explored only as proof of concept because the degree of the approximation depends on the interpolator which in turn will be impacted by smoothness and continuity of the manifold which is not guaranteed in a general sense for multi-modal problems.  

\subsection{Comparison with PT and Complexity}\label{sec:complex}

Parallel tempering with 5 chains was run alongside PT-STWNC on the model from section \ref{sec:SIR} so as to compare the algorithms.  The rule of thumb spacing for temperatures was used $\tau_t=(t/5)^5$ for $t\in\{1,\ldots,5\}$ following \citep{calderhead2009estimating}, although using fewer temperatures than would be required for low bias thermodynamic integration.  Both PT-STWNC and PT resulted in similar point and modal importance estimates with the latter values differing by at most 5\% across the 4 modes.  In both cases the lag 1 autocorrelation for the target chain was driven almost entirely by which mode it was sampling from and as a result in both methods that value was $\sim 0.97$ but differed slightly in the third decimal place.  

\begin{figure}[!h]
	\centerline{\includegraphics[scale=0.13]{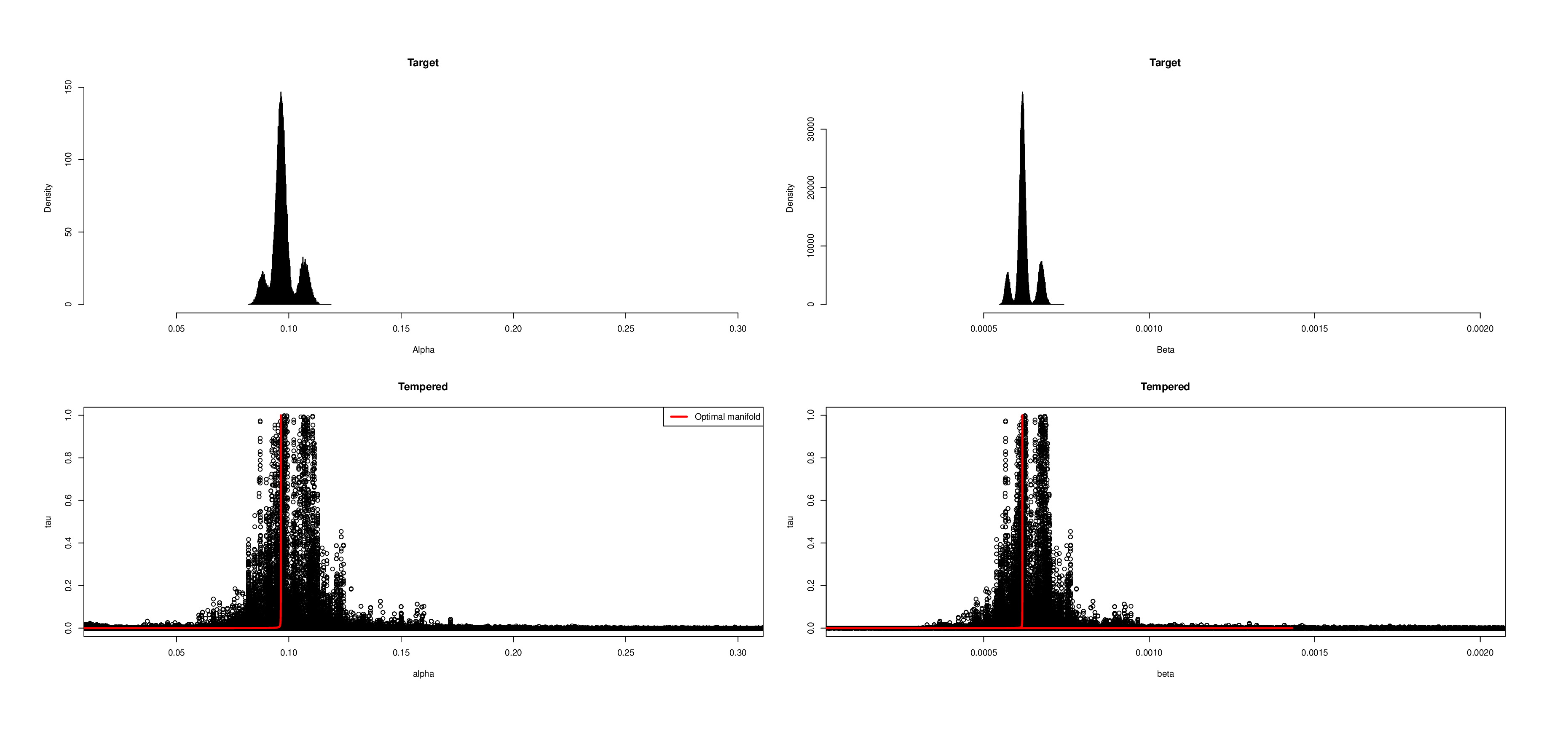}}
	\caption{Comparing the sampled values of $\beta$ and $\alpha$ with respect to $\tau$ along with the values found by the optimizer.}
	\label{fig:SIR2Dtau}
\end{figure}

When the optimum values are tracked, ST an additional diagnostic which is unavailable to PT in that posterior samples should be in the vicinity of the optimal value for all $\tau\in[0,1]$ if the chain is sampling correctly.  Figure \ref{fig:SIR2Dtau} shows the samples of $\beta$ and $\alpha$ plotted against $\tau$ for the STWNC chain.  This figure includes the optimum for the profile at each value of $\tau$.  The optimum line moves slowly with decreasing $\tau$ but curves dramatically near zero values as the priors (both exponential with mean 1) over-ride the likelihood.

\section{Discussion} \label{sec:Conclusions}

In this paper we presented a solution to the implementation challenge that has left Simulated Tempering unuseable as a  Monte Carlo method. Our approach allows ST to be used with a continuous temperature and without pilot runs or normalizing constant approximations.   STWNC is more computationally complex than PT but maintains the inherent capability of tempering algorithms to mix across challenging posterior regions.  Furthermore STWNC provides easy access to Bayes Factors through thermodynamic integration and provides an additional heuristic diagnostic of assessing sampling in the neighbourhood of the optima along the full range of $\tau$.  

Results from PT and PT-STWNC were comparable in all of our examples in terms of mixing of the target chain, estimation of points, and modal importance. However, STWNC is more computationally complex than PT because of the optimization stage.  In the way the SIR example was set up in section \ref{sec:SIR}, one would need to include 50 PT chains to produce approximately the same time per iteration as the 2 chain PT-STWNC.  Results will depend on the complexity of the model in question and at the other extreme, the 2 chain PT and 2 chain PT-STWNC were approximately equivalent in speed in section \ref{sec:Example1} because the optima are analytically available. Optimization of the posterior distribution with respect to the parameters for a fixed $\tau$ may be initially computationally expensive in complex models. However, the optimization time decays with the number of MCMC iterations if previous optimal values of $\theta$ are used to initialize the optimizer in the next Markov Chain iteration.  Further reduction in the cost as iterations progress could be obtained by replacing the optimizer with a function evaluation.  Such a replacement would make the STWNC chain as fast as a PT chain but with much more capacity to overcome distant modes.   The novelty of the STWNC is in the mathematical nature of the solution rather than its computational speed, but ongoing work shows considerable promise to accelerate the sampler.  Additionally, having an implementable ST method allows additional work on exploiting the geometry of the problem, producing alternative pathways between the prior and posterior, and exploiting the now obtainable $\tau$ distribution as part of a modified Hamiltonian Monte Carlo method.

\bigskip
\begin{center}
{\large\bf COMPUTER CODE}
\end{center}
Examples in this paper are implemented in \citep{Rlang}, using following packages: utils, graphics, parallel, deSolve \citep{deSolve}, MASS \citep{citeMASS}, gtools \citep{gtools}, MCMCpack \citep{MCMCpack}, mvtnorm \citep{mvtnorm}, truncnorm \citep{truncnorm}, optimx \citep{optimx}, coda \citep{coda}.
Relevant R code and data is provided in the public github repository 
https://github.com/BiljanaJSJ/PT-STWNC.


\bibliographystyle{spbasic}
\bibliography{ref}

\section{Apendices}
\subsection{Analytical calculation of the marginal likelihood} \label{sec:AppendixA}
In this section we provide the details on analytic calculation of the marginal likelihood in (\ref{eq:ordmllik}) for the bimodal model example given in the Section (\ref{sec:Example1}).

The posterior distribution of the unknown parameter $\mu$ is,
\small
\begin{eqnarray}
P(\mu \mid \sigma^{2},\boldsymbol{Y}) &\propto& P(\boldsymbol{Y} \mid  |\mu| , \sigma^{2} ) P(\mu) 
= \prod_{i=1}^{n}\text{N}(  y_{i} \mid |\mu|, \sigma^{2}) P(\mu)  \nonumber \\
&=&\left(\prod_{i=1}^{n}\text{N}(y_{i} \mid \mu, \sigma^{2}) \mathbb{I}(\mu>0)+\prod_{i=1}^{n}\text{N}(-y_{i} \mid \mu, \sigma^{2}) \mathbb{I}(\mu<0)\right)   \nonumber \\ & & \times P(\mu)  \nonumber \\
\end{eqnarray}

\normalsize
where the variance was fixed at $\sigma^{2}=1$. The prior of $\mu$ was Gaussian:
$P(\mu)  \sim \text{N}(\lambda=0,\beta=1)$.

The Thermodynamic Integral is: 

\begin{eqnarray}
\int\limits_{\mu} P(\mu  \mid \sigma^{2} , \boldsymbol{Y})d\mu  
&=&(2\pi)^{-\frac{n}{2}}(\sigma^{2})^{-\frac{n}{2}}\beta^{-1}\frac{1}{2}\sqrt{a^{-1}}   \nonumber \\
&\times& \left[ \exp{ \left( \frac{1}{2} \frac{ b^{2}}{   a      }   - \frac{1}{2} c \right) }  + 	\exp{ \left( \frac{1}{2} \frac{ b_{1}^{2}}{   a      }   - \frac{1}{2} c \right) }  \right], \nonumber \\
\label{eq:thInt}
\end{eqnarray}
where $a=\frac{n}{\sigma^{2}}+\frac{1}{\beta}$, $b= \frac{\sum_{i}y_{i}}{\sigma^{2}}+\frac{\lambda}{\beta}$, $b_{1}= \frac{-\sum_{i}y_{i}}{\sigma^{2}}+\frac{\lambda}{\beta}$ and $c= \frac{\sum_{i}y_{i}^{2} }{\sigma^{2}}  +\frac{\lambda^{2}}{\beta} $.

Plug in the values of $\{\boldsymbol{Y}, \lambda,\beta,\sigma^{2},n\}$ in the solution of the integral given by the equation (\ref{eq:thInt})  and take a log to obtain the analytical marginal likelihood reported in Table~{\ref{tbl:pants}}.

\subsection{Implementation, bimodal model} \label{ImplementDetails}

The transition kernel of $\mu$ was updated using the optimal symmetric jumping kernel for Gaussian target distributions by \citep{gelman1996efficient}. The proposal distribution for of $\mu$ at the the $i$-th iteration is:

\begin{equation} 
\mu^{(i+1)}  \sim \text{N} \left(\mu^{(i)}, [2.4/\sqrt{d}]^{2}   \mathrm{Var}_{P(\mu \mid \sigma^{2},\textbf{Y},\tau)}(\mu)  \right)  \label{ST:adapt}\end{equation}
where $\mathrm{Var}_{P(\mu \mid \sigma^{2},\textbf{Y},\tau)}(\mu)$ is the target variance of $\mu$ with respect to the target posterior distribution  $ P(\mu \mid \sigma^{2},\boldsymbol{Y},\tau)$, $[2.4/\sqrt{d}]^2$ is the optimal scale factor of the target variance found by \citep{gelman1996efficient} with d being the dimension of the parameters updated in the MCMC step. The variance parameter $\sigma^{2}$  was sampled from a log normal proposal distribution. The transition step was tuned so that the acceptance rate is 44 $\%$. The inverse temperature parameter was updated by drawing independent samples from the standard uniform proposal distribution.

In order to evaluate the prior $P(\tau)$ in (\ref{EQ:ptau}),  $\mu$ and $\sigma^{2}$ were optimized using closed forms of the conditional posterior mean of $P(\mu \mid \boldsymbol{Y},\sigma^{2},\tau)$ and the conditional posterior mode of $P(\sigma^{2} \mid \boldsymbol{Y}, \mu ,\tau)$, respectively. In particular, $\mu$ and $\sigma^2$ were maximized in a conditional iterative manner by optimizing each of them conditional on the last optimized value of the other. Iterations were repeated until the optimized values of the both parameters stopped changing within a tolerance level of $10^{-3}$. Using explicit formulae of posterior means (or modes) avoids numerical issues that are usually associated with optimization routines.

\paragraph*{Marginal likelihood estimation using thermodynamic integration via Parallel Tempering -- bimodal model}

Convergence of each of the PT chains used to obtain the TI-PT-NB and the TI-PT-B estimates was assessed  using the Potential Scale Reduction Factor (PSRF) or $\hat{\text{R}}$ statistics by  \citep{gelman1992inference}, which compares the in-chain and between-chain variances of the chains for each of the $20$ runs. The observed $\hat{R} <1.1$ in our runs indicates that the chains have converged.

\subsection{Implementation PT-STWNC, SIR model}  \label{ImplementSIR}

The proposal distribution of $\tau$ was a truncated standard normal at [0,1]. The proposal distributions of  $\alpha$ and $\beta$ were log-normal. The proposal distribution of the $I(0)$ was Binomial, set up as follows: the proposed at the $i-th$ iteration is: $\text{I}(0)^{(i)}=\mbox{Binomial}(\text{N},\frac{\text{I}(0)^{(i-1)}}{\text{N}})$, where $\text{N}=261$ is the total population.

Optimization of    $\boldsymbol{\theta}=\left(\alpha,\beta, \text{I}(0) \right)^{'}$, which is needed to evaluate the prior of the inverse temperature $\tau$,  was carried out by first optimizing the continuous $(\alpha,\beta)^{'}$ conditional on fixed discrete values of  $\text{I}(0)=\{1,..,8\}$ using the Nelder-Mead optimization routine. Then, out of the eight optimized values $\left(\alpha_{\max},\beta_{\max},I(0) \mid \text{I}(0) \  \in \{1,2,..,8\}\right)^{'}$, the one that maximizes the posterior distribution $P(\alpha,\beta,\text{I}(0) \mid  \boldsymbol{Y}, \tau)$ was chosen as a maximum.

\end{document}